\documentclass[twocolumn]{aastex631}

\shorttitle{Detecting Atmospheric $\mathrm{CO_2}$ Trends on Temperate Terrestrial Exoplanets}
\shortauthors{Hansen et al.}

\usepackage{amsmath}
\usepackage{enumitem}
\usepackage{hyperref}
\usepackage{bm}

\begin{document}

\title{Detecting Atmospheric $\mathrm{CO_2}$ Trends as Population-Level Signatures for Long-Term Stable Water Oceans and Biotic Activity on Temperate Terrestrial Exoplanets}

\correspondingauthor{Janina Hansen}
\email{jahansen@phys.ethz.ch}

\author[0009-0003-1247-8378]{Janina Hansen}
\affiliation{ETH Zurich, Institute for Particle Physics \& Astrophysics, Wolfgang-Pauli-Str. 27, 8093 Zurich, Switzerland}
\affiliation{National Centre of Competence in Research PlanetS (www.nccr-planets.ch)}

\author[0000-0001-6138-8633]{Daniel Angerhausen}
\affiliation{ETH Zurich, Institute for Particle Physics \& Astrophysics, Wolfgang-Pauli-Str. 27, 8093 Zurich, Switzerland}
\affiliation{National Centre of Competence in Research PlanetS (www.nccr-planets.ch)}
\affiliation{Blue Marble Space Institute of Science, Seattle, WA, USA}
\affiliation{SETI Institute, 189 N. Bernado Ave, Mountain View, CA 94043, USA}

\author[0000-0003-3829-7412]{Sascha P. Quanz}
\affiliation{ETH Zurich, Institute for Particle Physics \& Astrophysics, Wolfgang-Pauli-Str. 27, 8093 Zurich, Switzerland}
\affiliation{National Centre of Competence in Research PlanetS (www.nccr-planets.ch)}
\affiliation{ETH Zurich, Department of Earth and Planetary Sciences, Sonneggstrasse 5, 8092 Zurich, Switzerland}

\author[0000-0002-6140-6325]{Derek Vance}
\affiliation{ETH Zurich, Institute of Geochemistry and Petrology, Department of Earth and Planetary Sciences, Clausiusstrasse 25, 8092 Zurich, Switzerland}

\author[0000-0002-9912-8340]{Björn S. Konrad}
\affiliation{ETH Zurich, Institute for Particle Physics \& Astrophysics, Wolfgang-Pauli-Str. 27, 8093 Zurich, Switzerland}
\affiliation{National Centre of Competence in Research PlanetS (www.nccr-planets.ch)}

\author[0000-0003-2530-9330]{Emily O. Garvin}
\affiliation{ETH Zurich, Institute for Particle Physics \& Astrophysics, Wolfgang-Pauli-Str. 27, 8093 Zurich, Switzerland}
\affiliation{National Centre of Competence in Research PlanetS (www.nccr-planets.ch)}

\author[0000-0002-0006-1175]{Eleonora Alei}
\affiliation{NASA Postdoctoral Program Fellow, NASA Goddard Space Flight Center, 8800 Goddard Rd, Greenbelt, 20771, MD, USA}

\author[0000-0003-2769-0438]{Jens Kammerer}
\affiliation{European Southern Observatory, Karl-Schwarzschild-Straße 2, 85748 Garching, Germany}

\author[0000-0002-5476-2663]{Felix A. Dannert}
\affiliation{ETH Zurich, Institute for Particle Physics \& Astrophysics, Wolfgang-Pauli-Str. 27, 8093 Zurich, Switzerland}
\affiliation{National Centre of Competence in Research PlanetS (www.nccr-planets.ch)}

\begin{abstract}
Identifying key observables is essential for enhancing our knowledge of exoplanet habitability and biospheres, as well as improving future mission capabilities. While currently challenging, future observatories such as the Large Interferometer for Exoplanets (LIFE) will enable atmospheric observations of a diverse sample of temperate terrestrial worlds. Using thermal emission spectra that represent conventional predictions of atmospheric $\mathrm{CO_{2}}$ variability across the Habitable Zone (HZ), we assess the ability of the LIFE mission - as a specific concept for a future space-based interferometer - to detect $\mathrm{CO_{2}}$ trends indicative of the carbonate-silicate (Cb-Si) weathering feedback, a well-known habitability marker and potential biological tracer. Therefore, we explore the feasibility of differentiating between $\mathrm{CO_{2}}$ trends in biotic and abiotic planet populations. We create synthetic exoplanet populations based on geochemistry-climate predictions and perform retrievals on simulated thermal emission observations. The results demonstrate the robust detection of population-level $\mathrm{CO_{2}}$ trends in both biotic and abiotic scenarios for population sizes as small as 30 Exo-Earth Candidates (EECs) and the lowest assessed spectrum quality in terms of signal-to-noise ratio, $S/N$\,=\,10, and spectral resolution, $R$\,=\,50. However, biased $\mathrm{CO_{2}}$ partial pressure constraints hinder accurate differentiation between biotic and abiotic trends. If these biases were corrected, accurate differentiation could be achieved for populations with $\geq$\,100\,EECs. We conclude that LIFE can effectively enable population-level characterization of temperate terrestrial atmospheres and detect Cb-Si cycle driven $\mathrm{CO_{2}}$ trends as habitability indicators. Nevertheless, the identified biases underscore the importance of testing atmospheric characterization performance against the broad diversity expected for planetary populations.
\end{abstract}


\section{Introduction} \label{sec:intro}
Within just a few decades, the search for potentially habitable and inhabited exoplanets has evolved from science fiction to a central scientific pursuit for the exoplanet community. Recent advancements, from transit surveys such as the Kepler mission \citep{Borucki_2010} and the Transiting Exoplanet Survey Satellite (TESS, \citealt{Ricker_2015}) to the launch of the James Webb Space telescope (JWST, \citealt{Gardner_2023}), have shifted focus from exoplanet detection to characterization. However, even with JWST, characterizing temperate terrestrial exoplanet atmospheres remains challenging and time-intensive (e.g. \citealt{Morley_2017, Krissansen-Totton_2018, De_Wit_2024}).

In the near future, ground-based observatories such as the extremely large telescopes (ELTs) have the potential to provide further data on the closest terrestrial exoplanets in reflected stellar light and thermal emission (e.g. \citealt{Quanz_2015, Bowens_2021, Kasper_2021}). Still, comprehensive atmospheric characterization of a significant population of these objects remains elusive, even with current and under-construction observatories. This limitation has spurred the development of next-generation mission concepts like the Habitable Worlds Observatory (HWO) and the Large Interferometer for Exoplanets (LIFE), aimed at deciphering temperate terrestrial atmospheres for habitability and biological activity indicators in reflected light and thermal emission, respectively \citep{HWO_2021, Miller_2024, Kammerer&Quanz2018, Quanz_2022, Quanz_2022_LIFEI}. As we prepare for these future missions, the task of ensuring that their designs are able to reliably perform these characterizations is one that requires action now. 

Numerous potential signatures indicative of surface oceans and global biospheres have been proposed (e.g. \citealt{Meadows_2008, Kaltenegger_2010, Grenfell_2017, Schwieterman_2018, Schwieterman_2024}). While individual planet characterization typically requires extensive contextual evidence, future missions like LIFE will enable the characterization of temperate terrestrial atmospheres on a statistically relevant sample. This capability raises the question of what we might learn about planetary habitability and biospheres by constraining planetary and atmospheric parameters on a broader scale.

One potential answer lies in the well-established concept of the Habitable Zone (HZ, \citealt{Kasting1993, Kopparapu2013, Kopparapu2014}). The HZ describes an orbital range around a host star where temperature conditions could allow for liquid water on a terrestrial planetary surface. In the conservative HZ definition, a $\mathrm{H_{2}O}$-$\mathrm{CO_{2}}$ greenhouse sustains liquid surface water on an Earth-like planet due to the climate-mediating effect of the carbonate-silicate (Cb-Si) cycle \citep{Kasting1993}. On Earth, atmospheric $\mathrm{CO_{2}}$ is linked to surface temperature by the Cb-Si weathering feedback, hypothesized to have stabilized the Earth’s climate over geological timescales \citep{Walker1981, Sagan_1972, Tajika_1992, Tajika_1993}. 

Translating this thermostat concept to a population of HZ planets residing at different orbital distances leads to an expected decreasing trend in atmospheric $\mathrm{CO_{2}}$ with increasing incident stellar flux \citep{Kasting1993, Kopparapu2013, Kadoya_2014, Graham_2020}. Detecting the predicted trend in atmospheric $\mathrm{CO_{2}}$ abundance with incident stellar flux provides a unique observable directly linked to the long-term presence of liquid surface water, a fundamental prerequisite for the Cb-Si cycle to function \citep{Walker1981, Kasting1993}. Moreover, it is commonly acknowledged that biological factors, particularly land plants, influence the Cb-Si cycle by enhancing silicate weathering \citep{Arthur1993, Moulten2000,Ibarra_2019, Porder_2019, Dahl_2020}. By assessing the ability to distinguish between $\mathrm{CO_{2}}$ trends on worlds hosting large-scale biospheres and abiotic planet populations, we explore the observational feasibility of $\mathrm{CO_{2}}$ trends as population-level clues for global biotic activity.

Previous studies investigated the feasibility of observing atmospheric $\mathrm{CO_{2}}$ abundances to infer population-level trends. As a proof of concept, \citet{Bean_2017} utilized an idealized artificial planet population to propose that a decreasing $\mathrm{CO_{2}}$ trend with incident stellar flux could be identified in a sample of 20 exoplanets, assuming ±\,0.5\,dex uncertainty in $\mathrm{CO_{2}}$ abundance. The weathering feedback was only implicitly included by fixing the surface temperature over the HZ stellar irradiation range. \citet{Lehmer2020} employed a more complex coupled geochemical-climate model, confirming a decreasing trend of $\mathrm{CO_{2}}$ abundance with incident flux but demonstrating considerable scatter. They suggested that a sample of 83 planets would be required to detect the Cb-Si distribution, excluding observational uncertainties. Further investigations by \citet{checlair_thesis_2021} estimated that 17-34 planets would be necessary to detect the weathering feedback, incorporating simplified observational uncertainty estimates for the  HabEx \citep{Gaudi2020} and LUVOIR \citep{Peterson_2017, LUVOIR_Team_2019} mission concepts, though without a comprehensive geochemical model. More recently, \citet{Lustig-Yaeger_2022} introduced a Hierarchical Bayesian Atmospheric Retrieval (HBAR) framework for hypothesis testing on predicted population-scale atmospheric trends. The authors validated their model by simulating JWST transmission spectra of a synthetic population of TRAPPIST-1e-like planets, which followed the theoretical $\mathrm{CO_{2}}$ trend proposed by \citet{Bean_2017}. The authors were able to reject the null hypothesis, represented by a flat trend, as well as other simple atmospheric trend models at high confidence but acknowledged the challenge of resolving varying abundances of $\mathrm{CO_{2}}$ with JWST-like observations.

Beyond $\mathrm{CO_{2}}$ trends, other studies have explored the detectability of population-level signatures in synthetic exoplanet populations. The Bioverse framework, for example, combined survey simulations with Bayesian hypothesis testing to evaluate whether future transit and direct imaging missions could constrain statistical relationships, such as a correlation between planet age and atmospheric $\mathrm{O_2}$ abundance or a discontinuity in planetary radii at the inner edge of the HZ \citep{Bixel_2021, Schlecker_2024}. These studies, albeit focused on different observables, emphasized the importance of assessing mission concepts in terms of their ability to recover statistical trends across planetary populations.

\begin{figure*}[tb]
    \centering
    \includegraphics[width=\linewidth]{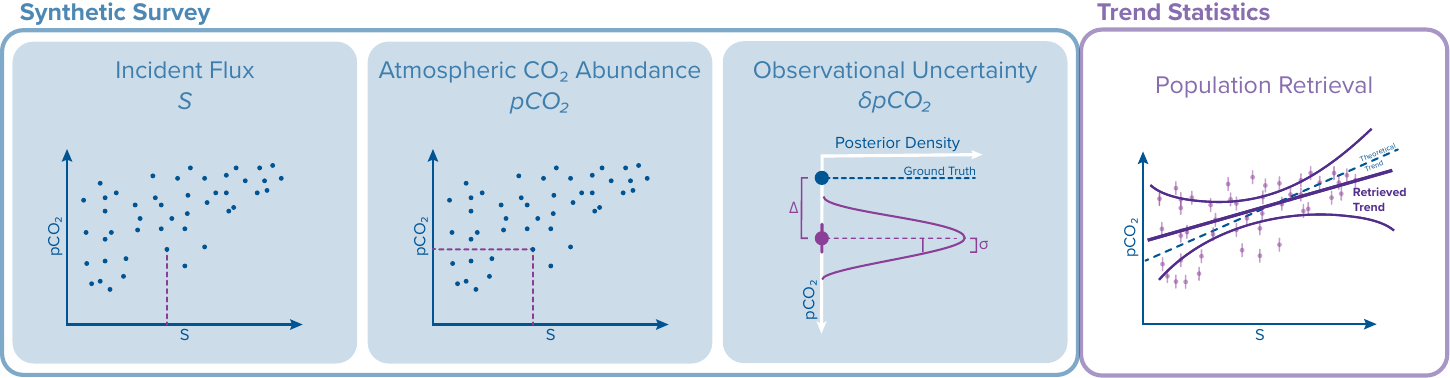}
    \caption{Schematic of our trend survey and inference methodology. We generate synthetic atmospheric populations by associating each planet with an incident flux $S$, injecting the predicted Cb-Si cycle-driven p$\mathrm{CO_{2}}$ variability across the HZ, and assigning observational uncertainties $\delta$p$\mathrm{CO_{2}}$. These populations then serve as input for the trend retrieval routine.}
    \label{fig:pipeline}
\end{figure*}

This present study explores how population-level trends in atmospheric $\mathrm{CO_{2}}$ abundance as a function of stellar irradiation can be robustly detected, as well as biotic and abiotic trends accurately differentiated, based on thermal emission spectra of temperate terrestrial exoplanet populations. We employ a comprehensive approach that combines detection yield simulations for a mid-infrared nulling interferometer, exemplified by the planned LIFE mission, self-consistent climate modeling throughout the HZ, and predictions for broad Cb-Si distributions based on geochemical-climate modeling results presented in \citet{Lehmer2020}. We perform atmospheric retrievals on a variety of emission spectra and implement a trend inference routine following the HBAR approach as reported in \citet{Lustig-Yaeger_2022}.

Our framework for creating synthetic survey populations is presented in Section \ref{sec:syn_survey}, followed by population-level atmospheric retrievals and trend analyses in Section \ref{sec:pop_retrievals}. Section \ref{sec:discussion} further explores the diagnostic potential of trend detection and differentiation, highlighting identified characterization biases and their impact on trend detection capabilities. Finally, we summarize our findings and offer insights into future research directions in Section \ref{sec:conclusions}.

\section{Synthetic Survey and Trend Inference Methods}\label{sec:syn_survey}
Synthetic exoplanet survey populations represent the foundation of our $\mathrm{CO_{2}}$ trend detection and differentiation analysis. To assess trend detection capabilities, the following three parameters need to be specified for each observational target:

\begin{itemize}
    \item $S$: incident flux received at the planet (Section \ref{subsec:flux_dist})

    \item p$\mathrm{CO_{2}}$: atmospheric $\mathrm{CO_{2}}$ partial pressure (Section \ref{subsec:co2_trend})

    \item $\delta$p$\mathrm{CO_{2}}$: observational uncertainty around a true p$\mathrm{CO_{2}}$ (Section \ref{subsec:uncertainty_statistics_II})
\end{itemize}

In the subsequent sections, we detail the process of assigning specific parameters to each planet within our synthetic survey samples. These survey populations then function as input to our trend inference routine (Section \ref{subsec:HBAR}), ultimately allowing us to assess detection capabilities. To provide a clear overview of this study's analytical steps, we have included a visual representation of our methodology in Figure \ref{fig:pipeline}.

\subsection{Incident Flux}\label{subsec:flux_dist}
To set up synthetic atmospheric populations which reflect a Cb-Si-driven correlation between $\mathrm{CO_{2}}$ abundance and stellar irradiation, we need to associate each planet with a level of incident flux $S$, which is characteristic for observable temperate terrestrial exoplanets residing in the HZ. This considers not only the underlying distribution of exoplanets across the HZ but also the observational biases that influence the perceived distribution. 

For this, we use a two-step process: First, we generate underlying exoplanet populations based on occurrence rate estimates from \citet{Bryson2021} using the population synthesis tool \texttt{P-Pop} \citep{Kammerer&Quanz2018}. Second, the mission simulator LIFE\textsc{sim} \citep{LIFE_II} is used to estimate the integration time required to detect each planet within the synthetic populations, providing us with a LIFE-characteristic HZ detection yield distribution. We outline the most important aspects of both steps in the following and refer to the original publications for further details \citep{Kammerer&Quanz2018, LIFE_II}.

\begin{figure}
    \centering
    \includegraphics[width=\linewidth]{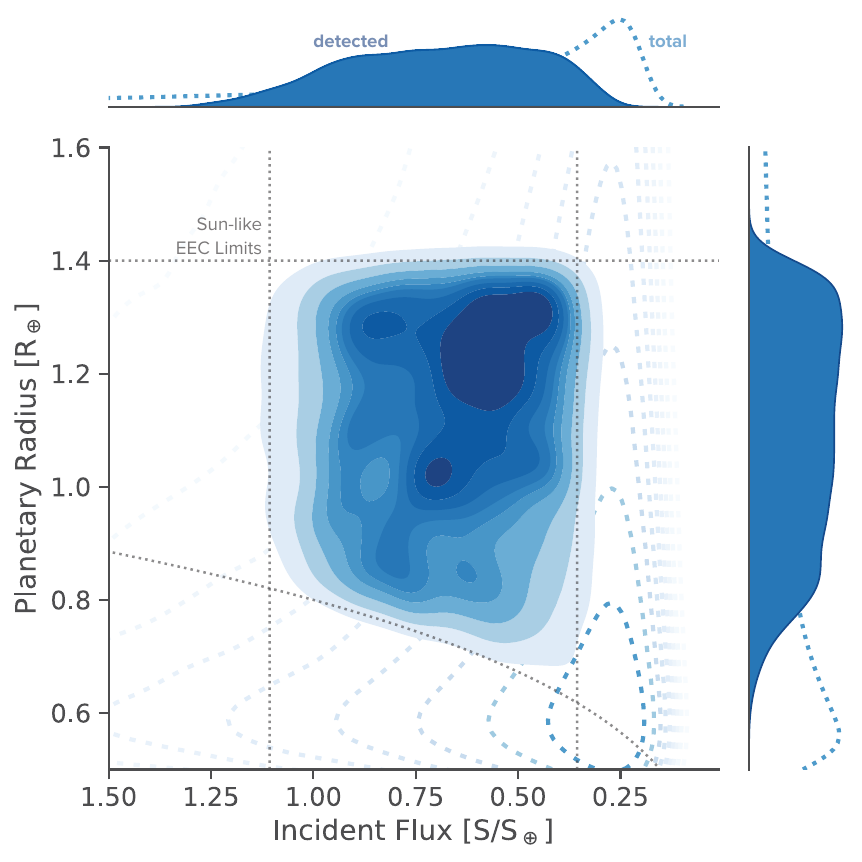}
    \caption{Distribution of simulated EEC yield (solid contours) and underlying planet population (dotted contours) around FGK-type stars within 20 pc. We show the planetary radius over incident flux distribution, where the color shading represents the relative density of planet occurrences in this parameter space. The colored contours enclose 10\% to 100\% of the total planet counts, with the innermost contour indicating the highest-density regions. Grey dashed lines represent EEC boundaries around a Sun-like star with $S/S_{\oplus}$\,$\in$\,[0.356, 1.107] and $R/R_{\oplus}$\,$\in$\,[0.8\,$\cdot$\,$S^{0.25}$, 1.4] \citep{Kopparapu2014, Stark2019, Quanz_2022_LIFEI}. Top and right margin plots show the individual distributions of incident flux and planetary radius of detected and total underlying populations. Underlying synthetic planet populations are based on \citet{LIFE_VI}.}
    \label{fig:flux_dist}
\end{figure}

To generate the underlying exoplanet population, we select FGK-type stars (3940-7220\,K effective temperature) within 20\,pc of the Sun from the LIFE target catalog \citep{Quanz_2022_LIFEI, Menti_2024}. With this, we aim to produce planet populations which remain close to an Earth-Sun-like environment. Each system is assigned an exozodiacal dust level from the nominal exozodi level distribution with a median of $\sim$\,3\,zodi \citep{Ertel2018, Ertel2020}. For each star in our target sample, we generate 1000 synthetic planet populations. The large number of Monte Carlo simulations enables marginalization over population properties, reflecting uncertainties in the occurrence rate model. The underlying planet population which we present here was generated by \citet{LIFE_VI} and further analyzed in \citet{Angerhausen_2024}. Additional details can be found in the respective publications. The total underlying population is visualized as dotted contours in Figure \ref{fig:flux_dist}.

To produce the HZ detection yield population, LIFE\textsc{sim} estimates the integration time required to detect each planet, optimizing the distribution of observational time for habitable zone planets. Here, we limit survey populations to exo-Earth candidates (EECs), the most conservative HZ planet category (see Figure \ref{fig:flux_dist} for EEC bounds after \citealt{Kopparapu2014,Stark2019, Quanz_2022_LIFEI}). We simulate photon noise from stellar leakage and zodiacal dust emission. Similar to previous studies (e.g. \citealt{LIFE_II, LIFE_VI, Angerhausen_2024}), we assume a conservatively high $S/N$ of 7 integrated over the full LIFE-wavelength range for detection, partially to account for omitted instrumental noise. Future LIFE\textsc{sim} implementations will incorporate a comprehensive treatment of instrumental noise (\citealt{Huber_2024}; Huber et al., in prep.; Dannert et al., in prep.). For each target star, we optimize the interferometer's baseline length to maximize transmission for objects located at the HZ center. We assume LIFE mission parameters of 2\,m aperture size, 5\% system throughput and 4-18.5\,$\mathrm{\mu m}$ wavelength coverage \citep{konrad2022}. Other specifications correspond to those outlined in \citet{Quanz_2022_LIFEI}. The derived EEC distribution is visualized as solid contours in Figure \ref{fig:flux_dist}.

We show the estimated distribution of detected EECs as well as the total underlying exoplanet population in the form of radius-flux distributions in Figure \ref{fig:flux_dist}. We observe a tendency to detect EECs toward the upper planetary radius limit, with a slight bias toward detecting planets in the outer incident flux range compared to the inner range. Comparing the derived yield distribution to the underlying population, we observe a detection bias toward larger radii, associated with enhanced thermal emission. We also note a detection preference for lower incident fluxes, which correlates with the maximum of the underlying population being located at the lower incident flux bound. Despite this, the derived yield distribution spans the entire incident flux range, providing broad access to survey planets in our trend detection context.  When comparing these results to modified mission implementations (such as varying aperture diameters), we do not observe significant changes in the incident flux distribution. While the derived EEC yield distribution corresponds to exoplanets orbiting FGK-type stars, Figure \ref{fig:flux_dist} includes Sun-like EEC bounds for reference. In subsequent stages of our synthetic survey setup, we constrain our analysis to Sun-like stellar irradiation bounds. However, this results in only a minor reduction in the accessible incident flux range, as $\sim$\,90\% of the detected EECs around FGK-type stars fall within these limits.

\subsection{Atmospheric \texorpdfstring{$\mathrm{CO_{2}}$}{CO2} Abundance}\label{subsec:co2_trend}
\begin{figure*}
    \centering
    \includegraphics[width=\linewidth]{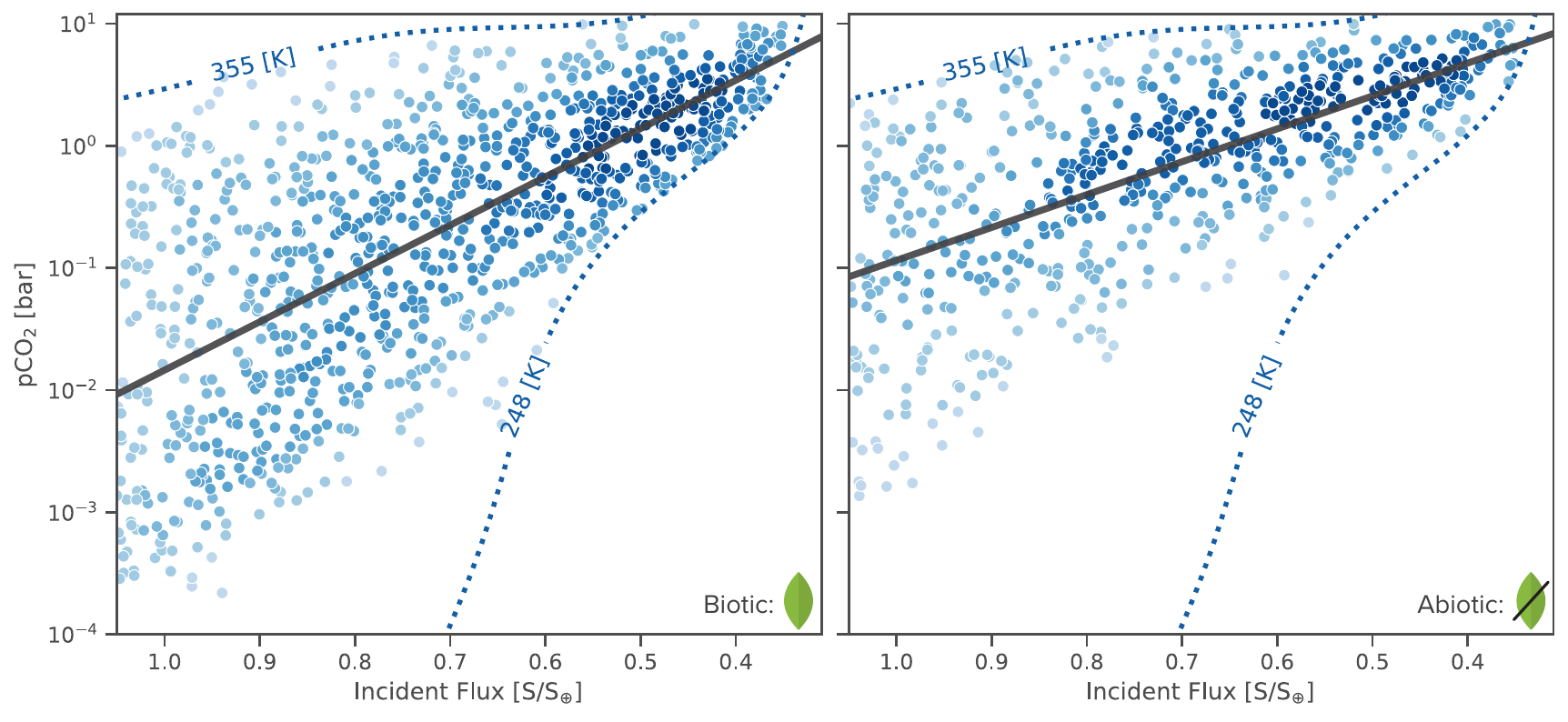}
    \caption{2D p$\mathrm{CO_{2}}$-$S$ distributions using coupled climate and Cb-Si cycle models \citep{Kriss-Tott2017, Kriss-Tott2018, Lehmer2020}. Left panel: Biotic distribution reproducing \citet{Lehmer2020}. Right panel: Abiotic distribution showing a limited biosphere contribution to the Cb-Si weathering feedback. Black lines represent weighted least squares (WLS) trend fits, yielding $\beta_{bio}$\,=\,3.96$\pm$0.18 (95\%) and $\beta_{abio}$\,=\,2.69$\pm$0.14 (95\%) with unit $\mathrm{-log_{10}(p\mathrm{CO_{2}}[bar])/(S/S_{\oplus})}$ for biotic and abiotic trend slopes, respectively. Our $\beta_{bio}$ slightly differs from \citet{Lehmer2020} due to correction for heteroskedasticity in the p$\mathrm{CO_{2}}$-$S$ distribution. The color gradient represents the density of data points. Temperature contours indicate surface temperature limits for stable liquid water on an Earth-like planet \citep{Charnay_2013,Wolf_2017}.}
    \label{fig:bio_abio_dist}
\end{figure*}

To inject the theoretically predicted p$\mathrm{CO_{2}}$-$S$ trend into our synthetic HZ planet populations, we use the coupled geochemistry-climate model introduced by \citet{Lehmer2020}. The authors applied the Cb-Si weathering framework reported in \citet{Kriss-Tott2017} and \citet{Kriss-Tott2018} to a population of Earth-like planets within Sun-like EEC incident flux bounds. \citet{Lehmer2020} assumed broad variations of parameters that influence the chemical weathering process, such as weathering rates, $\mathrm{CO_{2}}$ outgassing, and land fractions. Their results show a considerably scattered 2D p$\mathrm{CO_{2}}$-$S$ distribution following a decreasing semi-logarithmic trend of $\mathrm{CO_{2}}$ partial pressure with increasing incident flux. The left panel of Figure \ref{fig:bio_abio_dist} shows a recreation of the 2D p$\mathrm{CO_{2}}$-$S$ distribution from \citet{Lehmer2020}. For this, we use the authors' supplementary software \citep{Lehmer_Code} and assume parameter ranges which are identical to those used in \citet{Lehmer2020}.

On Earth, land plants contribute to the Cb-Si cycle by enhancing chemical weathering and increasing atmospheric $\mathrm{CO_{2}}$ drawdown (e.g., \citealt{Dahl_2020}). \citet{Lehmer2020} broadly parameterized biological weathering by uniformly sampling the biological weathering fraction ($f_{bio}$) over a range of 0.1 to 1.0, representing Archean and modern Earth weathering rates, respectively. The p$\mathrm{CO_{2}}$-$S$ distribution, referred to as biotic in this context, covers the parameter range introduced by \citet{Lehmer2020} and accounts for the assumed variability of biotically enhanced weathering rates from the early to the modern day Earth.

For this study, we additionally investigate the impact of a reduced biotic influence on the Cb-Si weathering feedback and the corresponding 2D p$\mathrm{CO_{2}}$-$S$ distribution. We limit the biological weathering fraction parameter range in the Cb-Si cycle model to $f_{bio}$ = [0.1, 0.25], representing predominantly abiotic, non-vegetated conditions. This reduction by a factor of four is based on observational studies comparing chemical weathering fluxes on vegetated versus non-vegetated stream catchments \citep{Arthur1993,Moulten2000}. A similar modification to the Cb-Si weathering model introduced in \citet{Kriss-Tott2017} and \citet{Kriss-Tott2018} was made by \citet{Hao_2020} to model the absence of vegetation in the Archean Earth. The reduced atmospheric $\mathrm{CO_{2}}$ drawdown shifts the abiotic p$\mathrm{CO_{2}}$-$S$ distribution toward larger $\mathrm{CO_{2}}$ partial pressures and higher surface temperatures (Figure \ref{fig:bio_abio_dist}, right panel). We use these predictions of biotic and abiotic distributions to associate planets in our synthetic survey population with incident flux-dependent $\mathrm{CO_{2}}$ partial pressures.

\subsection{Observational Uncertainty}\label{subsec:uncertainty_statistics_II}
In the final step of our synthetic survey setup, we assign an observational uncertainty ($\delta$p$\mathrm{CO_{2}}$) to each simulated planet, representing the error in constraining true atmospheric $\mathrm{CO_{2}}$ partial pressures. Preliminary hypothesis tests, detailed in Appendix \ref{sec:hypothesis_testing}, employed a simplified model of observational uncertainty, assuming uniform uncertainty across survey populations. These initial tests demonstrated that our ability to detect and differentiate trends is significantly influenced by the magnitude of observational uncertainty on p$\mathrm{CO_{2}}$.

For our primary trend detection analysis, we implement a more sophisticated approach to observational uncertainties, evaluating potential correlations between the constraints on retrieved p$\mathrm{CO_{2}}$ and a planet's position in the 2D p$\mathrm{CO_{2}}$-$S$ parameter space. We perform atmospheric retrievals on 12 atmospheric model scenarios representing characteristic, Cb-Si cycle injected, climates throughout the HZ. This approach provides more representative $\delta$p$\mathrm{CO_{2}}$ estimates for population-scale atmospheric diversity and enables direct translation of trend detection criteria to observational parameters such as signal-to-noise ratio, $S/N$, and spectral resolution, $R$.

We generate self-consistent atmospheric compositions using \texttt{ATMOS} \citep{arney2016,arney2017} and use \texttt{petitRADTRANS} \citep{molliere2019, Molliere_2020, alei2022} to generate atmospheric mid-infrared spectra. We then simulate observations with the LIFE\textsc{sim} tool and perform atmospheric retrievals using the framework introduced by \citet{konrad2022} and updated by \citet{alei2022} and \citet{konrad2023}. This framework uses \texttt{petitRADTRANS} as a forward model and \texttt{pyMultiNest} \citep{buchner2016} for parameter inference. Subsequent sections outline our approach to constructing atmospheric scenarios, generating simulated LIFE observations, and implementing the retrieval framework.

\subsubsection{Atmospheric Model Grid} \label{sec:atm_grid}
\begin{figure*}[htb]
    \centering
    \includegraphics[width=0.96\linewidth]{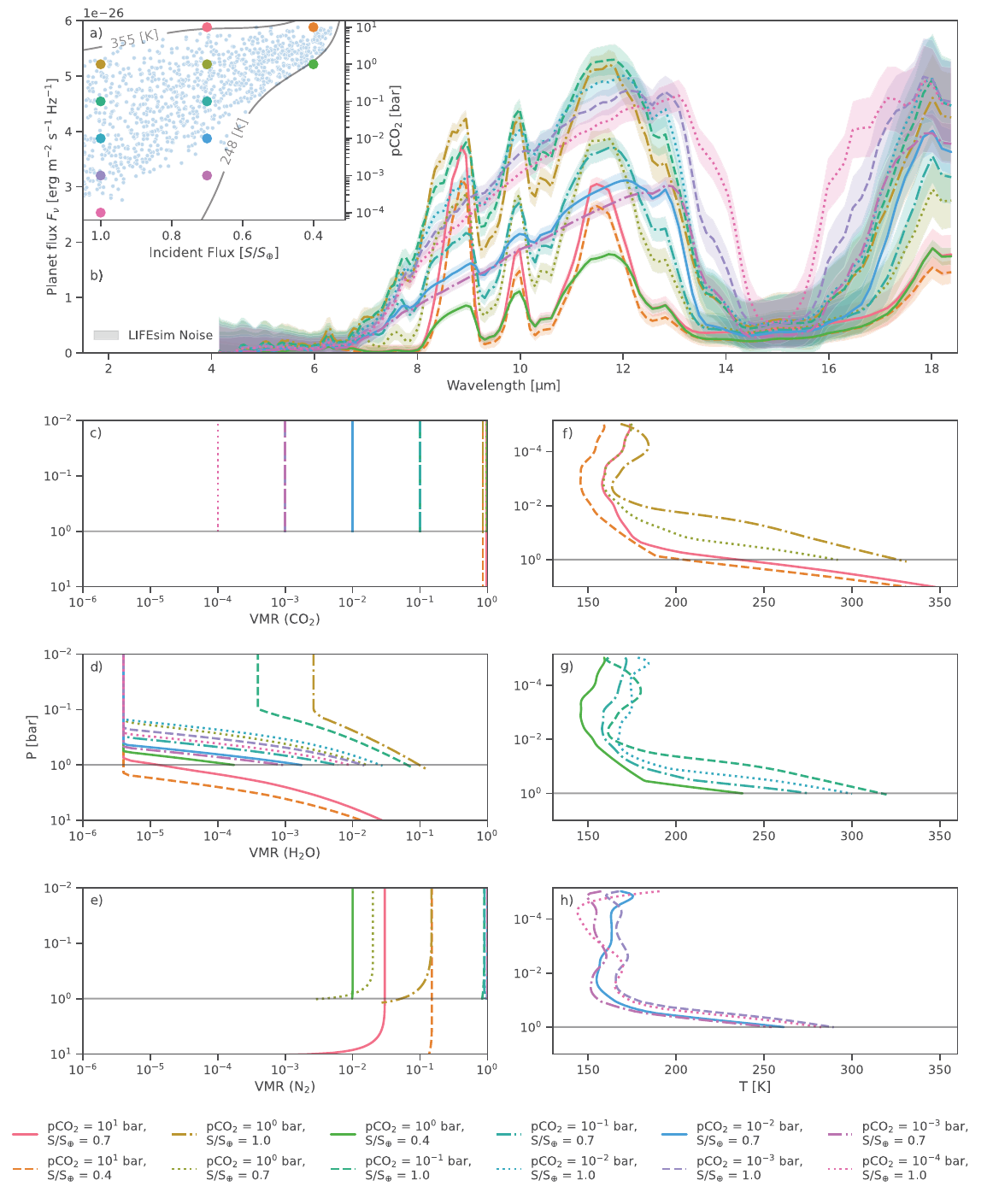}
    \caption{Atmospheric profiles and emission spectra for the 12 atmospheric scenarios covered by our grid. (a) Atmospheric grid structure visualization. (b) Simulated emission spectra with LIFE\textsc{sim} noise estimates for $S/N$\,=\,20 and $R$\,=\,100. (c)-(e) VMR profiles for $\mathrm{CO_2}$, $\mathrm{H_2O}$, and $\mathrm{N_2}$. (f)-(h) $P-T$ profiles grouped into three sets of four atmospheric scenarios each, to facilitate visualization. Color-coding in all panels corresponds to individual scenarios within the population.
    }
    \label{fig:spectra_species_profiles}
\end{figure*}

We create 12 self-consistent climate scenarios that represent the predicted variability of atmospheric p$\mathrm{CO_{2}}$ on Cb-Si planets within the HZ. For this purpose, we construct a grid in p$\mathrm{CO_{2}}$-$S$ space, encompassing the entire parameter range of the underlying planet population produced by \citet{Lehmer2020}. We model atmospheres with p$\mathrm{CO_{2}}$\,=\,[$10^{-4}$, $10^{-3}$, $10^{-2}$, $10^{-1}$, 1, 10]\,bar which reside at $S$\,=\,[0.4, 0.7, 1.0]\,$S_{\oplus}$ respectively. This discrete grid facilitates systematic comparison across various atmospheric scenarios. The atmospheric grid structure is visualized in Figure \ref{fig:spectra_species_profiles}. The ModernEarth CLIMA template \citep{arney2016,arney2017} serves as basis and is adapted according to the respective grid scenarios. In line with typical HZ climate simulations, our atmospheric compositions incorporate, besides $\mathrm{CO_{2}}$, $\mathrm{H_{2}O}$ as a primary greenhouse gas and $\mathrm{N_{2}}$ as a filler species. We assume constant $\mathrm{CO_{2}}$ volume mixing ratio (VMR) profiles. Pressure-dependent water vapor profiles are modeled self-consistently across the HZ within the CLIMA framework. For $\mathrm{H_{2}O}$, we employ the Manabe-Wetherald parametrization, featuring empirical tropospheric water vapor and constant stratospheric water content \citep{Manabe_1967}. $\mathrm{N_2}$ fills the remaining atmosphere such that the VMR in each atmospheric layer sums to unity. For scenarios with p$\mathrm{CO_{2}}$\,$\leq$\,1\,bar, we set the initial dry surface pressure to 1\,bar, while for p$\mathrm{CO_{2}}$\,=\,10 bar scenarios, it is set to 10\,bar. Figure \ref{fig:spectra_species_profiles} shows all generated abundance and pressure-temperature ($P-T$) profiles.

\subsubsection{Simulated LIFE Spectra}
\begin{table*}[htb]
{\footnotesize
\caption{Line and continuum opacities.}
\label{tab:opacities}
\begin{center}
        \setlength\extrarowheight{4pt} 
\begin{tabular}{cccccccc}
\hline\hline
\multicolumn{4}{c}{Molecular Line Opacities}&\multicolumn{2}{c}{CIA}&\multicolumn{2}{c}{Rayleigh Scattering}
\\
Molecule & Line List & Pressure Broadening & Wing cutoff & Pair & Reference & Molecule & Reference\\
\cline{1-4} \cline{5-6} \cline{7-8}
$\mathrm{CO_2}$ & HN20 & $\gamma_{air}$ & 25 $\mathrm{cm^{-1}}$ & $\mathrm{N_2}$-$\mathrm{N_2}$   & \citet{Karman_2019} & $\mathrm{N_2}$  & \citet{Thalman_2014, Thalman_2017}  \\
$\mathrm{H_2O}$ & HN20 & $\gamma_{air}$ & 25 $\mathrm{cm^{-1}}$ & $\mathrm{CO_2}$-$\mathrm{CO_2}$ & \citet{Karman_2019} & $\mathrm{CO_2}$ & \citet{Sneep_2005}  \\
$\mathrm{N_2}$  & HN20 & $\gamma_{air}$ & 25 $\mathrm{cm^{-1}}$ & $\mathrm{H_2O}$-$\mathrm{H_2O}$ & \citet{Karman_2019} & $\mathrm{H_2O}$ & \citet{Harvey_1998}  \\
                &      &                &                       & $\mathrm{H_2O}$-$\mathrm{N_2}$  & \citet{Karman_2019} &                 &   \\
\hline
\end{tabular}\end{center}}
\end{table*}

\begin{table}[tb]
{\footnotesize
\caption{Parameters of forward model and assumed priors in the retrieval.}
\label{tab:forward_model}
\begin{center}
        \setlength\extrarowheight{4pt} 
\begin{tabular}{lll}
\hline\hline
Parameter & Description & Prior \\
\hline
$a_4$           & $P$-$T$ parameter                 & $\mathcal{U}$(-10,10)    \\
$a_3$           & $P$-$T$ parameter                 & $\mathcal{U}$(-50,50)   \\
$a_2$           & $P$-$T$ parameter                 & $\mathcal{U}$(-100,500)   \\
$a_1$           & $P$-$T$ parameter                 & $\mathcal{U}$(0,500)   \\
$a_0$           & $P$-$T$ parameter                 & $\mathcal{U}$(0,1000)  \\
$\log_{10}$($P_0$)              & $\log_{10}$(Surface pressure [bar])           & $\mathcal{U}$(-4,3)    \\
$R_{pl}$                        & Planet radius [$R_{\oplus}$]                   & $\mathcal{G}$(1.0,0.2) \\
$\log_{10}$($M_{pl}$)             & $\log_{10}$(Planet mass [$M_{\oplus}$])        & $\mathcal{G}$(0.0,0.4) \\
$\log_{10}$($\mathrm{CO_2}$)    & $\log_{10}$($\mathrm{CO_2}$ mass fraction)    & $\mathcal{U}$(-15,0)* \\
$\log_{10}$($\mathrm{H_2O}$)    & $\log_{10}$($\mathrm{H_2O}$ mass fraction)    & $\mathcal{U}$(-15,0) \\
$\log_{10}$($\mathrm{N_2}$)     & $\log_{10}$($\mathrm{N_2}$ mass fraction)     & $\mathcal{U}$(-15,0)   \\
\hline
\end{tabular}
\tablecomments{$\mathcal{U}$($x$,$y$) indicates a uniform prior with a lower limit $x$ and upper limit $y$. $\mathcal{G}$($\mu$,$\sigma$) denotes a Gaussian prior with mean $\mu$ and standard deviation $\sigma$. *For p$\mathrm{CO_2}$\,$\geq$\,1\,bar cases, we apply uniform priors $\mathcal{U}$(0,1) on the $\mathrm{CO_2}$ mass fraction.}
\end{center}}

\end{table}

To generate noisy emission spectra that are representative of observations performed with LIFE, we follow a two-step approach: First, we generate mid-infrared emission spectra for every scenario in our atmospheric grid using the 1D plane-parallel radiative transfer code \texttt{petitRADTRANS} \citep{molliere2019, Molliere_2020, alei2022}. Second, we estimate the observational noise with the mission simulator tool LIFE\textsc{sim} \citep{LIFE_II}. While we use LIFE as specific mission concept, the subsequent analyses are based on defined spectrum quality metrics, such as $S/N$ and $R$, which are broadly applicable to comparable mission concepts.

We consider two signal-to-noise ratios $S/N$\,=\,[10,\,20] and two spectral resolutions $R$\,=\,[50,\,100] for all 12 spectra, based on preliminary LIFE requirements \citep{konrad2022,konrad2023,alei2022}. $R$ is defined as $\lambda / \Delta \lambda$, where $\lambda$ is the wavelength at the bin center and $\Delta \lambda$ denotes the bin width. Input for the radiative transfer routine includes pressure-dependent species profiles and $P-T$ profiles from the previous step (Section \ref{sec:atm_grid}). We use HITRAN 2020 (HN20; \citealt{Gordon_2022}) opacity line lists and account for collision-induced absorption (CIA) and Rayleigh scattering for all species (Table \ref{tab:opacities}). Cloud effects are omitted in our atmospheric models, with implications discussed in Section \ref{subsec:limitations}. We assume that all planets in our grid are at a 10\,pc system distance, orbiting a G2V star.

LIFE\textsc{sim} estimates wavelength-dependent noise to simulate LIFE observations of the spectrum population. Following previous analyses (e.g. \citealt{konrad2022, konrad2023, alei2022}), we define the $S/N$ of the spectrum as the $S/N$ in a reference wavelength bin. Here, we choose a reference bin at 11.5\,$\mathrm{\mu m}$, which is undisturbed by strong absorption features in our emission spectrum population. The $S/N$ at all other wavelengths is then determined by the LIFE\textsc{sim} noise model, which accounts for photon noise by the planet as well as major astrophysical noise sources. We do not randomize spectral points based on $S/N$, but understand $S/N$ as uncertainty around the true spectral points. Implications of this choice are discussed in detail in \citet{konrad2022}. Figure \ref{fig:spectra_species_profiles} displays the resulting modeled noisy emission spectra of our 12 atmospheric scenarios.

\subsubsection{Bayesian Atmospheric Retrieval Framework} \label{subsubsec:retrieval_framework}
With the help of atmospheric retrievals, observations of noisy spectra can be translated to estimates of planetary and atmospheric parameters. Retrieval routines typically consist of two components - a forward model and a parameter inference algorithm. The former infers atmospheric spectra on the basis of planetary and atmospheric parameters. The latter then searches the prior space that is defined by prior beliefs, prior probability distributions (priors), which we choose for the model parameters, and finds the parameter combinations that best fit the noisy spectrum we used as input to the retrieval. All possible parameter combinations are then provided in form of posterior probability distributions (posteriors) (for a review, see, e.g., \citealt{Madhusudhan_2018}).

Similar to the generation of our input spectra, we use \texttt{petitRADTRANS} to calculate theoretical 1D mid-infrared emission spectra as forward model in the retrieval. As parameter inference model we use \texttt{pyMultiNest} \citep{buchner2016}, a Python interface for the \texttt{MultiNest} nested sampling package \citep{Feroz_2009}. All retrievals were run with 600 live points and a sampling efficiency of 0.3, which is recommended for evidence evaluation. 

In the retrieval forward model we parameterize the $P-T$ profile with a fourth-order polynomial

\begin{equation}
    T(P) = \sum_{i=0}^4 a_i \log_{10}(P^i)\,.
\end{equation}

Here, $P$ represents pressure, $T$ the respective temperature and $a_i$ the parameters of the $P-T$ model. We choose a polynomial $P-T$ model in order to allow for a broad description of thermal structures, which is required in the context of analyzing a broad variety of atmospheric states within our atmospheric grid. We use the same line lists as in the input spectra and take into account CIA and Rayleigh scattering (see Table \ref{tab:opacities}). This choice is motivated by systematic biases that can be introduced to the retrieval when different sets of opacities are used in the input spectra and the retrieval forward model. For further discussions on this we refer to \citet{alei2022}. We assume constant abundance profiles for all atmospheric species and omit the effect of clouds in our retrieval framework. Implications of these simplifications are discussed in Section \ref{subsec:limitations}.

The priors we select for our model parameters are listed in Table \ref{tab:forward_model}. The priors on the $P-T$ profile parameters $a_i$ and the surface pressure $P_0$ are chosen to cover a wide range of thermal structures. We choose broad uniform prior ranges on the $\mathrm{\log_{10}}$ mass fraction for all species. In case of  the p$\mathrm{CO_2}$ \,$\geq$\,1\,bar scenarios in our atmospheric grid, we instead set a uniform prior on the mass fraction to enable an increased sampling density in the high abundance regime. This approach is justified as constraints around high $\mathrm{CO_2}$ abundances are retrieved when assuming log-uniform priors on the $\mathrm{CO_2}$ abundance. Such a result would naturally be followed by a retrieval inference allowing for increased sensitivity for high species abundances. Consistent with previous retrieval studies (see e.g. \citealt{konrad2022, konrad2023, alei2022}), we use Gaussian priors for planet radius $R_{pl}$ and mass $M_{pl}$. This choice is motivated by an expected constraint on $R_{pl}$ provided by the planet detection with LIFE \citep{LIFE_II}. Given the $R_{pl}$ prior, we can derive a prior on $\mathrm{log_{10}}$($M_{pl}$) by using the probabilistic mass-radius relation model \texttt{Forecaster} \citep{Chen_2017}.

\subsection{Population Trend Inference Routine} \label{subsec:HBAR}
To extract population-level atmospheric trends from a set of atmospheric observations and their corresponding retrieval estimates (i.e., p$\mathrm{CO_2}$ posteriors), we employ a hierarchical Bayesian atmospheric retrieval (HBAR) approach. This methodology, introduced by \citet{Lustig-Yaeger_2022}, is based on the importance sampling formalism developed by \citet{Hogg_2010}. In the following, we provide a description of the most relevant aspects of the HBAR framework and refer to the original publications for further details.

We perform a Bayesian inference on the population-level $\mathrm{CO_2}$ trend based on a series of simulated atmospheric observations. We introduce a set of hierarchical (hyper-) parameters $\boldsymbol{\gamma}$, which characterize the trend model. We then solve the Bayesian inverse problem for these parameters. Applying Bayes theorem we obtain

\begin{equation}
    \mathcal{P}(\boldsymbol{{\gamma}}\,\vline\,\{\textbf{D}_n\}_{n=1}^{N_P}) \propto \mathcal{L}_{\gamma}\,\mathcal{P}(\boldsymbol{\gamma}),
\end{equation} 

where $\mathcal{P}(\boldsymbol{{\gamma}}\,\vline\,\{\textbf{D}_n\}_{n=1}^{N_P})$ denotes the posterior for the population parameters $\boldsymbol{\gamma}$. $\mathcal{P}(\boldsymbol{\gamma})$ corresponds to the priors we set for the population parameters and $\mathcal{L}_{\gamma}$ is the likelihood. The latter is where $\mathrm{CO_2}$ abundance posteriors from individual planet observations come into play. Here, $\textbf{D}_n$ represents a set of spectroscopic observations for each planet $n$ in a total population of $N_P$ exoplanets. Given the population parameters $\boldsymbol{\gamma}$, the probability of the population-level dataset is then defined as

\begin{equation}
    \mathcal{L_{\gamma}} \equiv \mathcal{P}(\{\textbf{D}_n\}_{n=1}^{N_P}\vline\,\boldsymbol{{\gamma}})\,.
\end{equation}

After a set of rearrangements for which we refer to \citet{Lustig-Yaeger_2022}, we obtain the following expression

\begin{equation} \label{eq:likelihood}
    \mathcal{L_{\gamma}} \approx \prod_{n=1}^{N_P} \frac{1}{K} \sum_{k=1}^K \frac{\mathcal{P}_{\gamma}(f_{\mathrm{CO_2}nk})}{\mathcal{P}_0(f_{\mathrm{CO_2}nk})}.
\end{equation}

Equation \ref{eq:likelihood} considers $K$ posterior samples from the original atmospheric retrieval posteriors, where $k$ represents the sample index. $\mathcal{P}_0(f_{\mathrm{CO_2}nk})$ denotes the original uninformative prior for $\mathrm{CO_2}$ abundance used in the individual atmospheric retrievals. We introduce an updated prior probability distribution, $\mathcal{P}_{\gamma}(f_{\mathrm{CO_2}nk})$, which provides a more informed prior on $\mathrm{CO_2}$ abundance compared to the original uninformative priors. The posterior distributions, derived from the atmospheric retrieval grid, are implicitly included within the $K$ samples.

Until this point, we have not specified the population parameters $\boldsymbol{\gamma}$ that we intend to retrieve. For this, we set up a functional form for the theoretical population-trend. Similar to \citet{Lehmer2020}, we fit a semi-logarithmic trend to the data, modeling the partial pressure of $\mathrm{CO_2}$ as a function of incident flux. The functional form follows

\begin{equation}
    \mathrm{log_{10}}\,(\mathrm{pCO_2}) = f_{\mathrm{CO_2}}(\alpha,\beta) = \alpha + \beta \cdot S/S_{\oplus}\,,
\end{equation}

where $\alpha$ represents the log-intercept and $\beta$ corresponds to the logarithmic slope. With the trend model fixed, we give a description of the updated prior $\mathcal{P}_{\gamma}(f_{\mathrm{CO_2}nk})$ which corresponds to a normal distribution that is dependent on population parameters $\boldsymbol{\gamma}$

\begin{equation}
    \mathcal{P}_{\gamma}(f_{\mathrm{CO_2}nk}) = \mathcal{N}(f_{\mathrm{CO_2}nk} - f_{\mathrm{CO_2}}(\alpha,\beta), \sigma_{\mathcal{N}})\,.
\end{equation} 

Here, $\sigma_{\mathcal{N}}$ corresponds to the standard deviation of the normal distribution, which assigns high probability to p$\mathrm{CO_2}$ values of the theoretical trend lying near the retrieved p$\mathrm{CO_2}$ posterior samples. Finally, we introduce our population parameters $\boldsymbol{\gamma}$\,=\,[$\alpha$, $\beta$, $\sigma_{\mathcal{N}}$], for which we choose broad prior distributions

\begin{equation}
    \mathcal{P}(\boldsymbol{\gamma})  \begin{cases}
                                        \,\alpha \sim \mathcal{U}(-10,10) \\
                                        \,\beta \sim \mathcal{U}(-10,10) \\
                                        \,\sigma_{\mathcal{N}} \sim \mathcal{N}_{1/2}(0,1)\,.
                                      \end{cases}
\end{equation}

Following \citet{Lustig-Yaeger_2022}, we adopt the same prior for $\sigma_{\mathcal{N}}$, where $\mathcal{N}_{1/2}(0,1)$ denotes a half-normal distribution with a mean of 0 and a standard deviation of 1. The inferred posteriors of population parameters $\boldsymbol{\gamma}$ will then provide us with population-level trend constraints that encode the information from individually retrieved p$\mathrm{CO_2}$ posteriors.

\section{Trend Statistics: Population Retrieval Results}\label{sec:pop_retrievals}
We outline our retrieval results for the simulated atmospheric grid in Section \ref{subsec:results_retrievals}, discuss the setup of posterior maps in Section \ref{subsec:results_postmaps}, and show results of the population-level trend retrievals in Section \ref{subsec:results_hbar}.

\subsection{Atmospheric Retrieval Results}\label{subsec:results_retrievals}
\begin{figure*}
    \centering
    \includegraphics[width=0.84\linewidth]{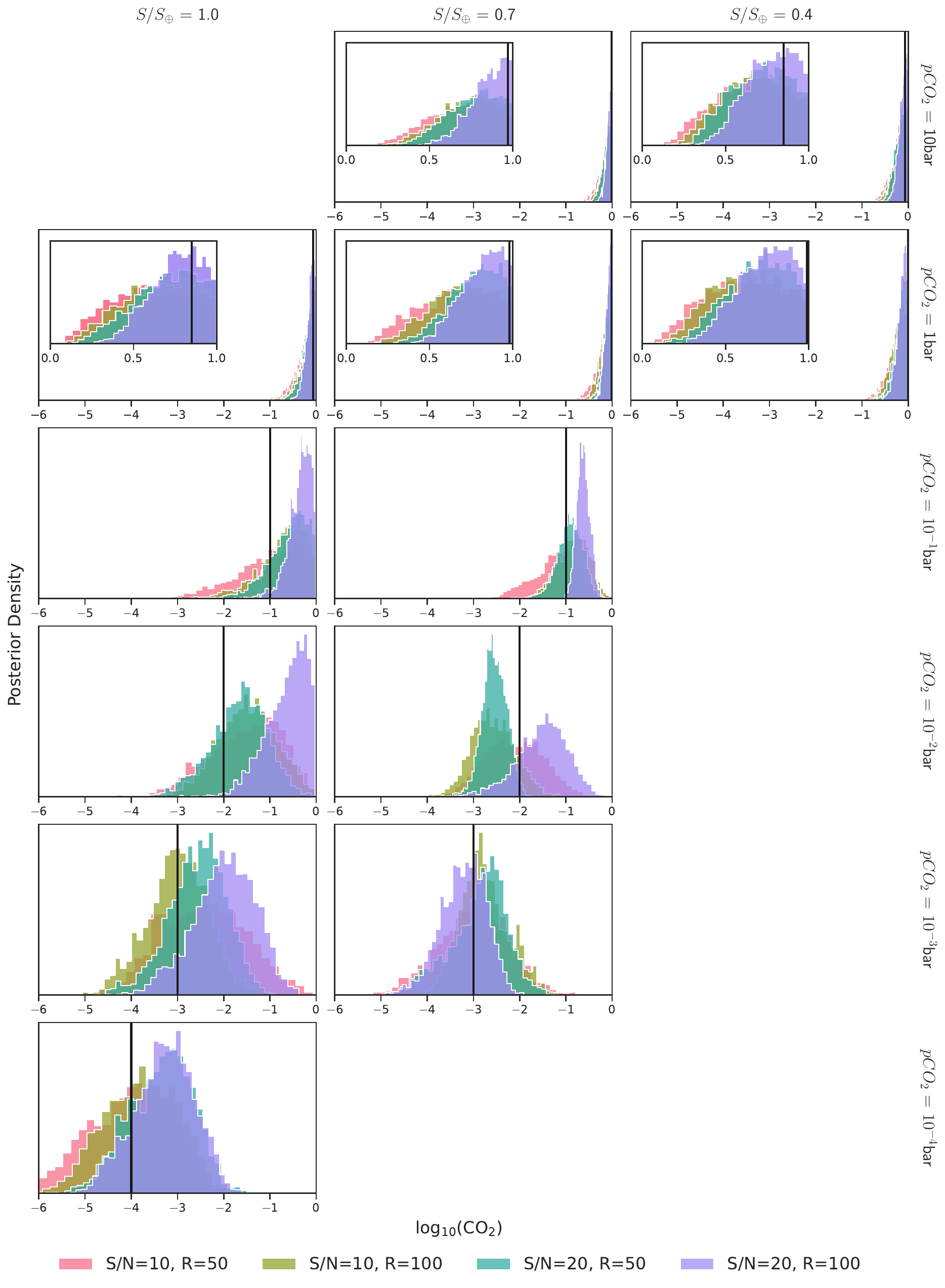}
    \caption{Posterior distributions of $\mathrm{CO_2}$ abundance derived from atmospheric spectrum retrievals. The panels are arranged to reflect the atmospheric grid within the p$\mathrm{CO_2}$-$S$ parameter space. Columns represent different stellar insolation values, $S/S_{\oplus}$\,=\,1.0, 0.7, 0.4, while rows correspond to different $\mathrm{CO_2}$ partial pressures, p$\mathrm{CO_2}$\,=\,10\,bar, 1\,bar, $10^{-1}$\,bar, $10^{-2}$\,bar, $10^{-3}$\,bar, $10^{-4}$\,bar. Each panel shows results for four combinations of signal-to-noise ratio, $S/N$\,=\,[10,\,20], and spectral resolution, $R$\,=\,[50,\,100]. Thick black lines indicate the ground truth $\mathrm{CO_2}$ abundance. For scenarios with p$\mathrm{CO_2}$\,$\geq$\,1\,bar, additional posterior distributions are plotted in non-logarithmic scale.
    }
    \label{fig:CO2_post}
\end{figure*}

We present atmospheric retrieval results for our grid of 12 atmospheric scenarios representing the predicted variability of $\mathrm{CO_2}$ abundance on Cb-Si planets across the HZ. In Figure \ref{fig:CO2_post} we show retrieved constraints on the $\mathrm{CO_2}$ abundance, i.e. volume mixing ratios (VMR), for all considered spectrum quality scenarios covering $S/N$\,=\,[10,\,20] and $R$\,=\,[50,\,100]. Moving forward, we use $\mathrm{CO_2}$ partial pressures to distinguish between different parameter spaces within the atmospheric population. This notation is chosen because p$\mathrm{CO_2}$, when combined with incident flux, serves as a unique identifier for our atmospheric scenarios, unlike $\mathrm{CO_2}$ VMRs. Additionally, we make a clear distinction between abundances (expressed as VMRs) and partial pressures.

Several qualitative observations can be made from Figure \ref{fig:CO2_post}. As the true atmospheric $\mathrm{CO_2}$ abundance increases, the constraints on the retrieved $\mathrm{CO_2}$ abundance become tighter. Higher observational sensitivity, characterized by increased $S/N$ and $R$, generally improves the precision (reduces standard deviations) of the retrieved posterior distributions. For most atmospheric scenarios and observational sensitivity cases, the $\mathrm{CO_2}$ abundance is well-constrained, with the ground truth falling within the 1$\sigma$ envelope of the retrieved parameter estimate. However, the accuracy (offset between ground truth and posterior median) of the constraints varies between p$\mathrm{CO_2}$\,$\geq$\,1\,bar and p$\mathrm{CO_2}$\,$<$\,1\,bar parameter spaces.

For p$\mathrm{CO_2}$\,$\geq$\,1\,bar scenarios, we additionally plot posterior distributions in non-log scale, since we chose non-log uniform priors to achieve a better sampling density for these high abundance cases. In this parameter space, we observe tight posterior constraints at the upper edge of the prior space, constraining well the ground truth $\mathrm{CO_2}$ abundances for all scenarios. For the p$\mathrm{CO_2}$\,$<$\,1\,bar cases,  we generally observe a tendency for $\mathrm{CO_2}$ posteriors to be offset toward larger abundances compared to the ground truth. For the highest observational sensitivity ($S/N$\,=\,20, $R$\,=\,100), the $\mathrm{CO_2}$ abundance is overestimated on average by a $\sim$\,1\,dex offset between ground truth and posterior median. With increasing true $\mathrm{CO_2}$ abundance, the bias in retrieved abundance estimates becomes more significant. As a result, the ground truth does not fall within the 1$\sigma$ posterior envelope for any observational sensitivity case in the p$\mathrm{CO_2}$\,=\,$10^{-1}$\,bar and $S/S_{\oplus}$\,=\,1.0 scenario. Moreover, the tendency to overestimate $\mathrm{CO_2}$ abundances is visibly more pronounced for atmospheres receiving an incident flux of $S/S_{\oplus}$\,=\,1.0 compared to $S/S_{\oplus}$\,=\,0.7 cases.

\subsection{\texorpdfstring{$CO_2$}{CO2} Posterior Maps}\label{subsec:results_postmaps}
Our atmospheric retrieval grid provides us with a population of 12 $\mathrm{CO_2}$ abundance posteriors for each spectrum quality scenario. To ensure consistency with the underlying theoretical p$\mathrm{CO_2}$-$S$ distributions and respective population trends, we translate abundance (VMR) posteriors to partial pressure posteriors via point-wise multiplication of retrieved $\mathrm{CO_2}$ abundance and surface pressure, $P_0$, constraints

\begin{equation}
    \mathcal{P}(\mathrm{pCO_2}) = \mathcal{P}(\mathrm{VMR}(\mathrm{CO_2})) \cdot \mathcal{P}(\mathrm{P_0}))\,, 
\end{equation}

where $\mathcal{P}$ represents the retrieved posterior. Subsequently, we generate characteristic posteriors maps for each spectrum quality scenario by interpolating between the p$\mathrm{CO_2}$ posteriors obtained from the atmospheric retrieval population. These posterior maps then allow us to associate any planet in our survey populations with a p$\mathrm{CO_2}$ posterior that is informed by our grid of atmospheric retrievals. For this, we fit Gaussian distributions to the derived p$\mathrm{CO_2}$ posteriors and then interpolate between obtained standard deviations $\sigma$ and offsets $\Delta$ (difference between posterior median and true partial pressures). In cases where the posterior distribution deviates from a well-defined Gaussian shape and instead provides only upper or lower bounds on p$\mathrm{CO_2}$, resembling a Gaussian peak with non-negligible tail on one side, our analysis focuses solely on the Gaussian component, which we refer to as the sensitivity peak. Following classifications in \citet{konrad2022}, we denote such posteriors as sensitivity-limited (SL).

\begin{figure}[tb]
    \centering
    \includegraphics[width=\linewidth]{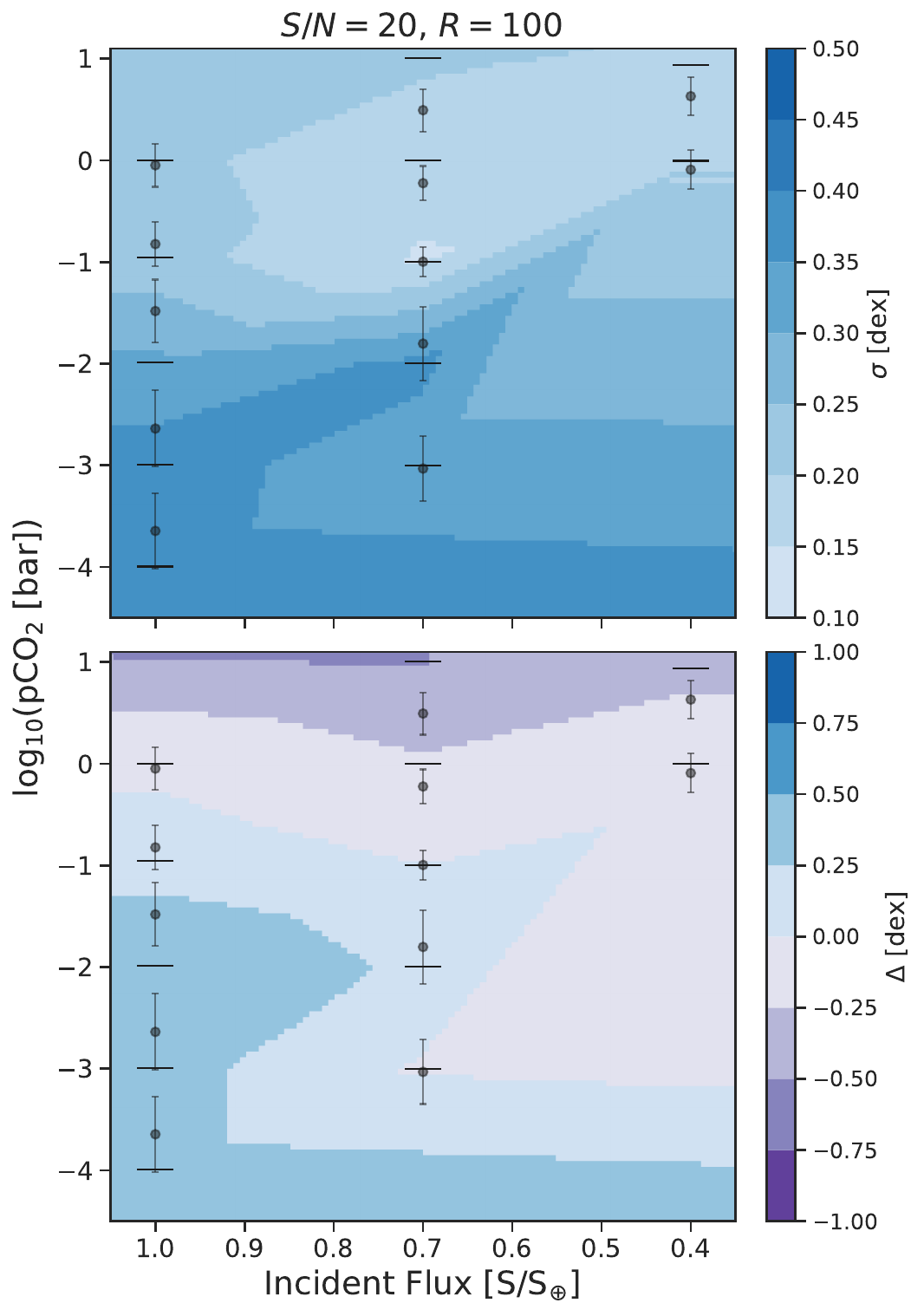}
    \caption{Maps of posterior standard deviations ($\sigma$) and offsets ($\Delta$) for observational sensitivity case $S/N$ = 20 and $R$ = 100. Color shading indicates the magnitude of $\sigma$ (top) and $\Delta$ (bottom), with blue and purple representing positive and negative offsets, respectively. Black horizontal lines denote ground truth $\mathrm{CO_2}$ partial pressures. Data points with error bars represent median and 1$\sigma$ envelope of Gaussian fits to retrieved p$\mathrm{CO_2}$ posterior distributions.
    }
    \label{fig:sigma_delta_map}
\end{figure}

We perform a linear interpolation between grid points situated at the ground truth p$\mathrm{CO_2}$ values. To achieve completeness, the maps are then expanded to cover the entire 2D p$\mathrm{CO_2}$-$S$ parameter space using nearest-neighbor interpolation. Hence, $\sigma$ and $\Delta$ maps may not accurately represent parameter spaces that fall substantially outside the established grid boundaries. Figure \ref{fig:sigma_delta_map} visualizes obtained $\sigma$ and $\Delta$ maps for the $S/N$\,=\,20 and $R$\,=\,100 scenario. A positive (negative) offset implies a shift of posteriors to higher (lower) partial pressures with respect to the true p$\mathrm{CO_2}$. Posterior maps generated for all spectral quality cases and the full population of retrieved p$\mathrm{CO_2}$ posteriors are shown in Appendix \ref{sec:app_supp_retrievals}. 

Similar to the abundance constraints (Section \ref{subsec:results_retrievals}), we identify wider posteriors toward low p$\mathrm{CO_2}$ and high $S/S_{\oplus}$ parameter ranges. However, we retrieve p$\mathrm{CO_2}$ posteriors which are narrower constrained and more centered around the ground truths ($\sigma$\,$\leq$\,0.4 dex, $\Delta$\,$\leq$\,0.75 dex). Systematic differences between the retrieved abundance and partial pressure constraints result from multiplying posteriors of $\mathrm{CO_2}$ abundance with surface pressure, two parameters that exhibit degeneracy in our retrievals. This will be discussed further in Section \ref{subsec:obs_biases}. For atmospheres with p$\mathrm{CO_2}$\,$\geq$\,1\,bar, we observe a trend of negative offsets in p$\mathrm{CO_2}$ posteriors relative to ground truths. In particular, scenarios with p$\mathrm{CO_2}$\,=\,10\,bar are underestimated, with ground truths falling outside the 1$\sigma$ posterior envelope. Conversely, in the p$\mathrm{CO_2}$\,$<$\,1\,bar range, we note a tendency for positive offsets between posterior means and true partial pressures.

\subsection{Trend Retrieval Results}\label{subsec:results_hbar}
We draw four sets of survey populations with $N_P$\,=\,[10, 30, 50, 100] planets. Based on the results of our preliminary hypothesis tests (see Appendix \ref{sec:hypothesis_testing}), which estimated the necessary sample sizes for detecting and distinguishing $\mathrm{CO_2}$ trends using a simplified observational uncertainty model, we choose a broad range of survey population sizes. For each population, we assign incident fluxes (Section \ref{subsec:flux_dist}), $\mathrm{CO_2}$ partial pressures (Section \ref{subsec:co2_trend}), and p$\mathrm{CO_2}$ posteriors corresponding to the respective location of the scenario in $\sigma$ and $\Delta$ maps (Section \ref{subsec:results_postmaps}).

\begin{figure*}[!h]
    \centering
    \includegraphics[width=\linewidth]{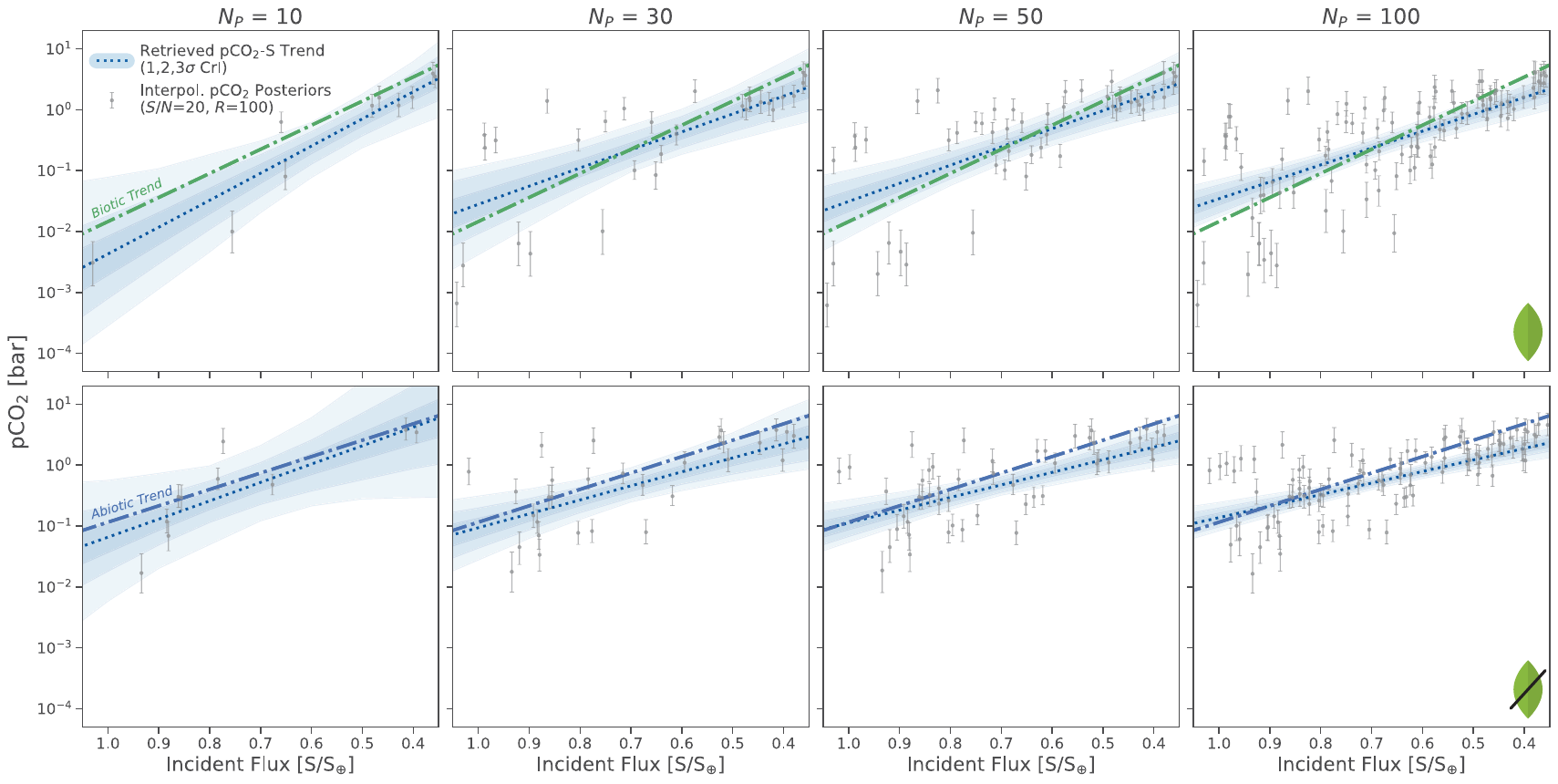}
    \caption{Retrieved population-level trends for biotic (top) and abiotic (bottom) scenarios for spectrum quality case $S/N$\,=\,20 and $R$\,=\,100. We differentiate between population sizes $N_P$\,=\,[10,\,30,\,50,\,100]. Dotted blue lines indicate median retrieved trends while blue shading indicates 1$\sigma$, 2$\sigma$ and 3$\sigma$ CrIs. Dashed lines represent theoretical biotic (green) and abiotic (blue) $\mathrm{CO_2}$ trends. Grey data points with error bars show median and 1$\sigma$ envelope of p$\mathrm{CO_2}$ posteriors, which correspond to a (p$\mathrm{CO_2}$, S) pair's location in $\sigma$ and $\Delta$ maps.}
    \label{fig:HBAR_trend}
\end{figure*}

\begin{figure*}[!h]
    \centering
    \includegraphics[width=\linewidth]{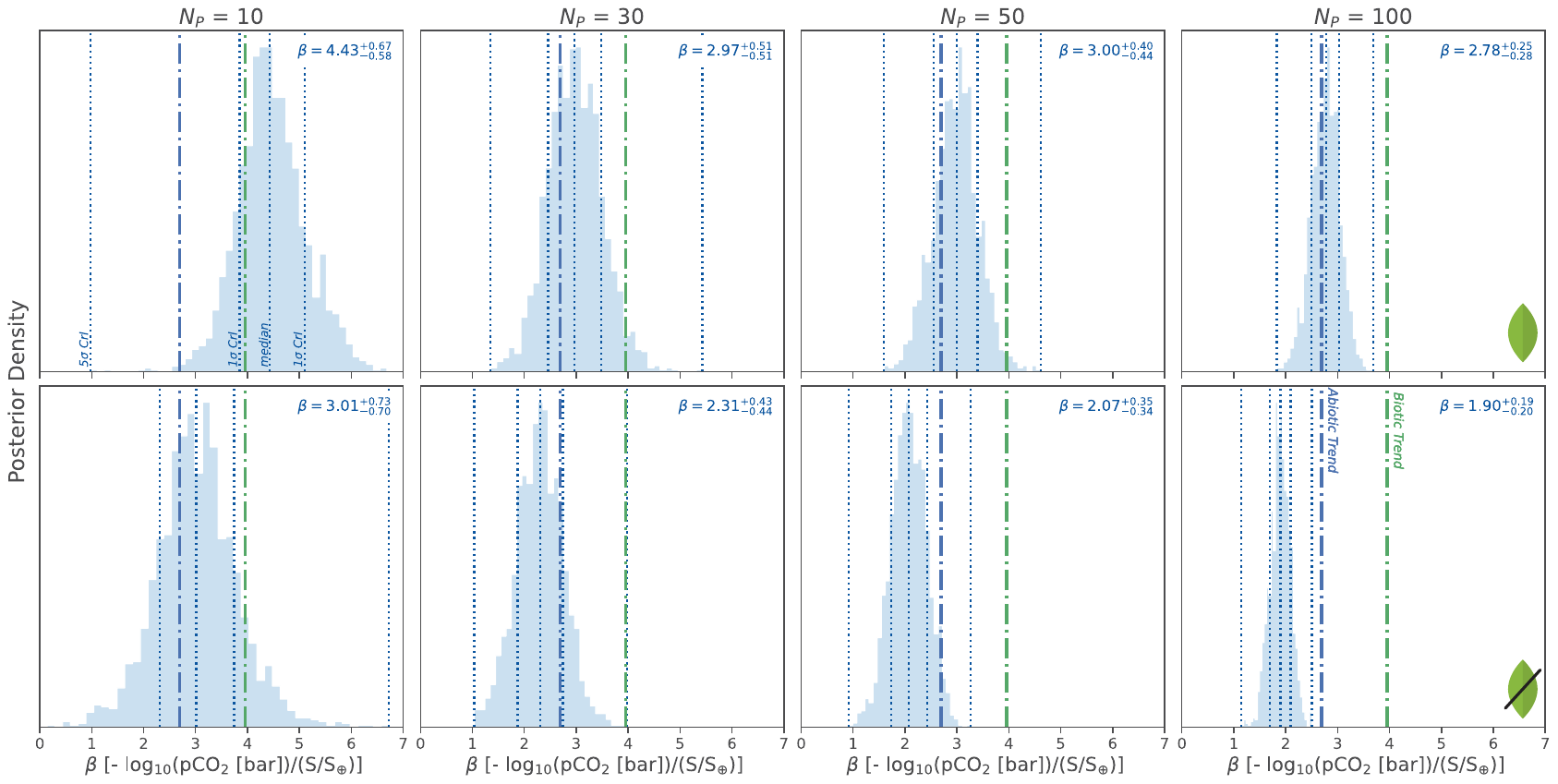}
    \caption{Retrieved posterior distributions of the slope parameter $\beta$ for biotic (top) and abiotic (bottom) scenarios for spectrum quality case $S/N$\,=\,20 and $R$\,=\,100. We differentiate between population sizes $N_P$\,=\,[10,\,30,\,50,\,100]. Dashed vertical lines indicate theoretical biotic (green) and abiotic (blue) $\mathrm{CO_2}$ trend slopes. Dotted blue lines represent median, 1$\sigma$, and 5$\sigma$ CrIs of the trend slope posterior. Median and 1$\sigma$ values of the retrieved posteriors are displayed in the top right for each scenario.}
    \label{fig:HBAR_posterior}
\end{figure*}

Figure \ref{fig:HBAR_trend} displays the retrieved population-level p$\mathrm{CO_2}$-$S$ trends for all $N_P$, encompassing both biotic and abiotic scenarios in the highest spectral quality case ($S/N$\,=\,20, $R$\,=\,100). The blue shaded areas indicate 1$\sigma$, 2$\sigma$, and 3$\sigma$ credible intervals (CrI) of the retrieved trends. Mirroring the underlying Cb-Si distributions (Section \ref{subsec:co2_trend}), p$\mathrm{CO_2}$ draws exhibit greater scatter at higher incident fluxes, particularly evident in high $N_P$ scenarios. The abiotic survey population shows less spread compared to the biotic one, primarily occupying the p$\mathrm{CO_2}$\,$\geq$\,$10^{-2}$\,bar parameter space. Comparing the drawn survey populations to the underlying Cb-Si distributions (see Figure \ref{fig:bio_abio_dist}) reveals populations which are more compressed along the p$\mathrm{CO_2}$ axis, reflecting the impact of offsets between associated posteriors and true p$\mathrm{CO_2}$ values. In line with the $\sigma$ map for the observational sensitivity case $S/N$\,=\,20 and $R$\,=\,100, broader posteriors of up to $\sim$\,0.4\,dex are observed for p$\mathrm{CO_2}$\,$<$\,1\,bar scenarios compared to p$\mathrm{CO_2}$\,$\geq$\,1\,bar cases.

Several general observations can be made: As expected, larger survey populations $N_P$ result in more precise constraints on the retrieved trend slope. However, these scenarios also demonstrate a more evident deviation of the retrieved trend constraint from the injected population-level trend. These patterns are also evident in the retrieved posteriors of slope parameter $\beta$, which are displayed for all $N_P$ and both biotic and abiotic scenarios in Figure \ref{fig:HBAR_posterior}. In general, the biotic scenario produces steeper trends compared to the abiotic one. The abiotic case yields more constrained slope estimates, which is consistent with a less dispersed underlying Cb-Si distribution.

For a survey population of $N_P$\,=\,10 planets in the biotic case, the null slope ($\beta$\,=\,0) falls beyond the 5$\sigma$ CrI of the retrieved trend slope posterior, indicating that a flat trend is highly improbable given the model and data. As $N_P$ increases to 30 or more, an even clearer distinction emerges between retrieved and flat trends for both biotic and abiotic cases. However, as already shown in Figure \ref{fig:HBAR_trend}, larger $N_P$ values lead to a more pronounced bias in trend detection, favoring flatter $\mathrm{CO_2}$ trends compared to the true injected population-level trend. For $N_P$\,=\,100, the true trend slopes lie outside the 5$\sigma$ CrI of the posterior distributions. Notably, the scenario characterized by the lowest observational sensitivity ($S/N$\,=\,10, $R$\,=\,50) does not show significant variations in the derived trend constraints, as depicted in Appendix \ref{sec:app_supp_trend_retrievals}.

\section{Discussion}\label{sec:discussion}
In the subsequent sections, we further evaluate the detection potential of $\mathrm{CO_2}$ trends in terrestrial HZ atmospheres (Section \ref{subsec:detection_potential}), drawing on our population-trend retrieval results, and address observational biases as well as their implications for trend detection capabilities (Section \ref{subsec:obs_biases}). Finally, we outline the limitations of this work in Section \ref{subsec:limitations}.

\subsection{Diagnostic Potential of population-level \texorpdfstring{$\mathrm{CO_2}$}{CO2} Trend Detection and Distinction}\label{subsec:detection_potential}
In Section \ref{subsec:results_hbar}, we inferred population-level $\mathrm{CO_2}$ trends based on synthetic survey populations and conducted an initial comparison of retrieved trend estimates with ground truth (biotic and abiotic) population trends, as well as flat trends representative of a non-existent relationship between p$\mathrm{CO_2}$ and the incident flux received by the planets. The following sections provide a deeper look at the detection and distinction potential of population-level $\mathrm{CO_2}$ trends.

\subsubsection{Detecting \texorpdfstring{$\mathrm{CO_2}$}{CO2} Trends}
To quantify the sensitivity of a $\mathrm{CO_2}$ trend detection (i.e. flat trend rejection) to population size $N_P$ and spectrum quality represented by $S/N$ and $R$, we use the Bayesian evidence $\mathcal{Z}$. $\mathcal{Z}$ is a model performance metric that, in this case, estimates how well the forward model fits the input trend population. By running our population-trend retrieval routine with two different trend models, a linear trend (Model A) and a flat trend (Model B, $\mathrm{log_{10}}\,(\mathrm{pCO_2}) = f_{\mathrm{CO_2}}(\alpha) = \alpha$), our pipeline yields different evidences dependent on the model. With these, we derive the Bayes' factor $K$:

\begin{equation}
    \mathrm{log}_{10}(K)\,=\,\frac{\ln(\mathcal{Z}_A) - \ln(\mathcal{Z}_B)}{\ln(10)}\,.
\end{equation}

The Bayes' factor indicates the preferred model. A positive $\mathrm{log}_{10}(K)$ indicates a preference for a linear trend (Model A) and a negative value shows a preference for a flat trend (Model B). The strength of this preference is further quantified on the basis of the Jeffreys scale (Appendix \ref{sec:app_supp_trend_retrievals}, \citealt{Jeffreys_1939}).

\begin{figure}[tb]
    \centering
    \includegraphics[width=\linewidth]{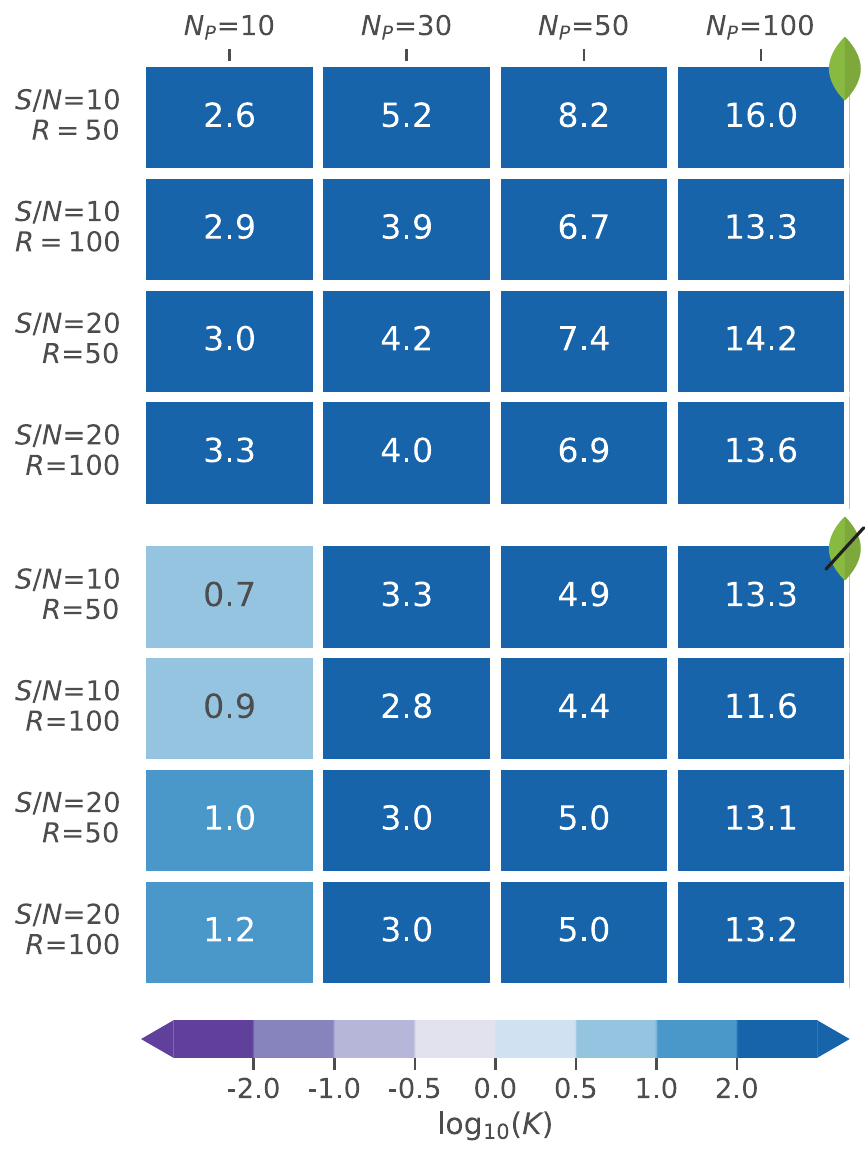}
    \caption{Bayes' factor comparison of linear and flat trend models for biotic (top) and abiotic (bottom) scenarios and for various combinations of population size $N_P$\,=\,[10,30,50,100] and spectrum quality , $S/N$\,=\,[10,20] and $R$\,=\,[50,100]. Positive $\mathrm{log}_{10}(K)$ values (blue) indicate a preference for the linear trend model, whereas negative $\mathrm{log}_{10}(K)$ values (purple) favor the flat trend model. The color intensity corresponds to the strength of preference according to the Jeffreys scale (Appendix \ref{sec:app_supp_trend_retrievals}, \citealt{Jeffreys_1939}).}
    \label{fig:Bayes_Factor}
\end{figure}

In Figure \ref{fig:Bayes_Factor}, we compare the Bayes' factor $K$ for all combinations of population size, $N_P$\,=\,[10,\,30,\,50,\,100], and spectrum quality, $S/N$\,=\,[10,\,20] and $R$\,=\,[50,\,100], assessed in this study. The top grid shows $\mathrm{log}_{10}(K)$ values for the biotic case, whereas the bottom grid indicates the abiotic scenario. In the biotic case, we see a decisive preference ($\mathrm{log_{10}}(K)$\,$>$\,2) for the linear over the flat trend model, which we denote here as a robust trend detection, for all population sizes and spectrum quality scenarios. In the abiotic $N_P$\,=\,10 case, we infer substantial (0.5\,$<$\,$\mathrm{log_{10}}(K)$\,$>$\,1) and strong (1\,$<$\,$\mathrm{log_{10}}(K)$\,$>$\,2) support for a linear model, in $S/N$\,=\,10 and $S/N$\,=\,20 cases respectively. For $N_P$\,$\geq$\,30, the linear model is again decisively preferred over the flat trend model.

In summary, the linear model is consistently preferred over the flat model. For $N_P$\,$\geq$\,30, a population-level $\mathrm{CO_2}$ trend is robustly detected for both biotic and abiotic cases in all regarded spectrum quality scenarios. Additionally, we identify a strong sensitivity of trend preference strength to $N_P$, while spectrum quality metrics $S/N$ and $R$ play a minor role in detection confidence. Yet, here we assess the diagnostic potential of one drawn population per respective $N_P$ scenario. Due to the wide spread of the assumed underlying Cb-Si distributions, a follow-up draw in the low $N_P$ regime might lead to differences in precision and accuracy of the retrieved population-level trend and, hence, might impact trend detectability for the lowest $N_P$ considered here.

\subsubsection{Differentiating Biotic and Abiotic \texorpdfstring{$\mathrm{CO_2}$}{CO2} Trends}
In Section \ref{subsec:results_hbar} we identified a biased retrieval of population-level trends toward flatter slopes for both biotic and abiotic cases, which become more pronounced with increasing planet population size. Here, we discuss these offsets and assess which factors play a role for precision and accuracy of our population-trend inferences.

\begin{figure}[tb]
    \centering
    \includegraphics[width=\linewidth]{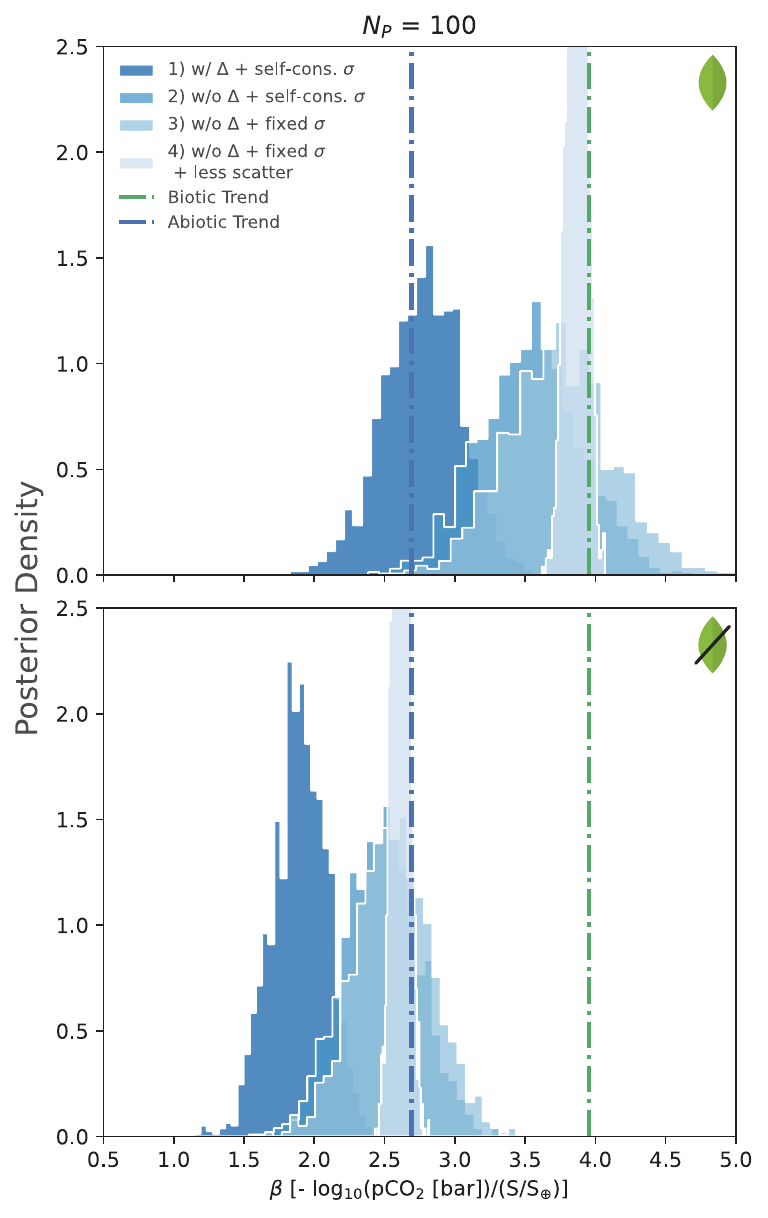}
    \caption{Comparison of trend distinction performance across four model assumptions in biotic (top) and abiotic (bottom) scenarios: 1) Base scenario, 2) No offset scenario, 3) Fixed standard deviation scenario, and 4) Reduced scatter scenario. We limit the model comparison to population size $N_P$\,=\,100 and spectrum quality $S/N$\,=\,20 and $R$\,=\,100.}
    \label{fig:Model_Comp}
\end{figure}

To evaluate the sensitivity of retrieved trend constraints to various parameters, we examine four distinct model assumptions within our trend retrieval analysis framework. Figure \ref{fig:Model_Comp} illustrates the trend slope posteriors for each model assumption scenario. In this analysis, we focus solely on the highest population size ($N_P$\,=\,100) and the most sensitive observational case ($S/N$\,=\,20, $R$\,=\,100), as these conditions allow for the clearest identification of differences in trend constraints.

The model comparison involves four scenarios to assess the ability to distinguish between biotic and abiotic $\mathrm{CO_2}$ trends:

\begin{enumerate}[label=\arabic*)]
    \item Base scenario: $\mathrm{CO_2}$ posterior distributions with standard deviations and offsets based on $\sigma$ and $\Delta$ maps, respectively.
    \item No offset scenario: Neglected offsets $\Delta$ between posterior means and true p$\mathrm{CO_2}$, with self-consistent posterior $\sigma$ for each planet.
    \item Fixed standard deviation scenario: Neglected offsets $\Delta$ and fixed posterior standard deviation of $\sigma$\,=\,0.1\,dex for all atmospheres.
    \item Reduced scatter scenario: Neglected offsets $\Delta$, fixed $\sigma$\,=\,0.1\,dex, and true p$\mathrm{CO_2}$ within ±0.1\,dex of the true $\mathrm{CO_2}$ trend.
\end{enumerate}

Our analysis of the four scenarios reveals distinct patterns in retrieved trend posteriors. In the first scenario, which corresponds to our base analysis setup, a notable shift toward flatter trends is observed for both biotic and abiotic cases. The second scenario shows trend posteriors which successfully constrain the ground truth within $\sim$\,$1\sigma$ CrI for both cases. A slight improvement in trend constraint accuracy is evident in the third scenario. The final scenario, which mimics a population with reduced scatter in p$\mathrm{CO_2}$, shows a significant increase in trend constraint precision.  

The comparison indicates that the posterior offset $\Delta$ significantly affects trend constraint accuracy, whereas the precision of trend constraints is largely influenced by population scatter, in addition to the planet population size $N_P$ as discussed in Section \ref{subsec:results_hbar}. This suggests that if offsets were to be corrected, an accurate trend retrieval could be achieved for a population size of $N_P$\,$\geq$\,100. Although observers cannot control the inherent variability of true p$\mathrm{CO_2}$ in an actual atmospheric population, future mission designs have the potential to optimize the number of planets in the population ($N_P$) and address biased p$\mathrm{CO_2}$ estimates ($\Delta$) derived from atmospheric retrieval frameworks.

Population sizes of this magnitude inevitably raise questions regarding feasible yields. While the LIFE mission progresses in its concept phase, previous studies have estimated potential exoplanet yields achievable with current implementation concepts (e.g., \citealt{Quanz_2022_LIFEI, LIFE_VI}). \citet{LIFE_VI} determined that a LIFE setup targeting FGK-type stars and comprising 3.5\,m mirrors and a total throughput of 5\% is expected to detect $20^{+20}_{-11}$ EECs, assuming optimistic exoplanet occurrence rates. By extending the boundaries of potentially habitable planets to the optimistic HZ definition according to \citet{Kopparapu2014}, the yield increases to $51^{+50}_{-27}$ planets. Comparable yields could be achieved with ${\scriptstyle \gtrsim}$\,2\,m mirrors and a throughput of 20\%. Although challenging, the $N_P$ requirements we have identified for $\mathrm{CO_2}$ trend detection and differentiation fall within the 1$\sigma$ uncertainties of yield predictions in the respective scenarios. While a science case as this could inform critical instrumentation design, it similarly drives the advancement of robust atmospheric characterization frameworks.

\subsection{Characterization Biases}\label{subsec:obs_biases}
We saw above that biases in parameter estimates provided by our atmospheric retrieval pipeline contribute to biased population-level trend inferences. While these play a minor role in trend detection, observational biases impact the ability to correctly differentiate biotic from abiotic $\mathrm{CO_2}$ trends. We discuss potential reasons for these p$\mathrm{CO_2}$ constraint biases in Sections \ref{subsec: H2O} and \ref{subsec:hab_limit}. In Section \ref{subsec:trend_strat} we discuss the implications of these observational biases for a trend detection strategy.

\begin{figure*}[tbp]
    \centering
    \includegraphics[width=\linewidth]{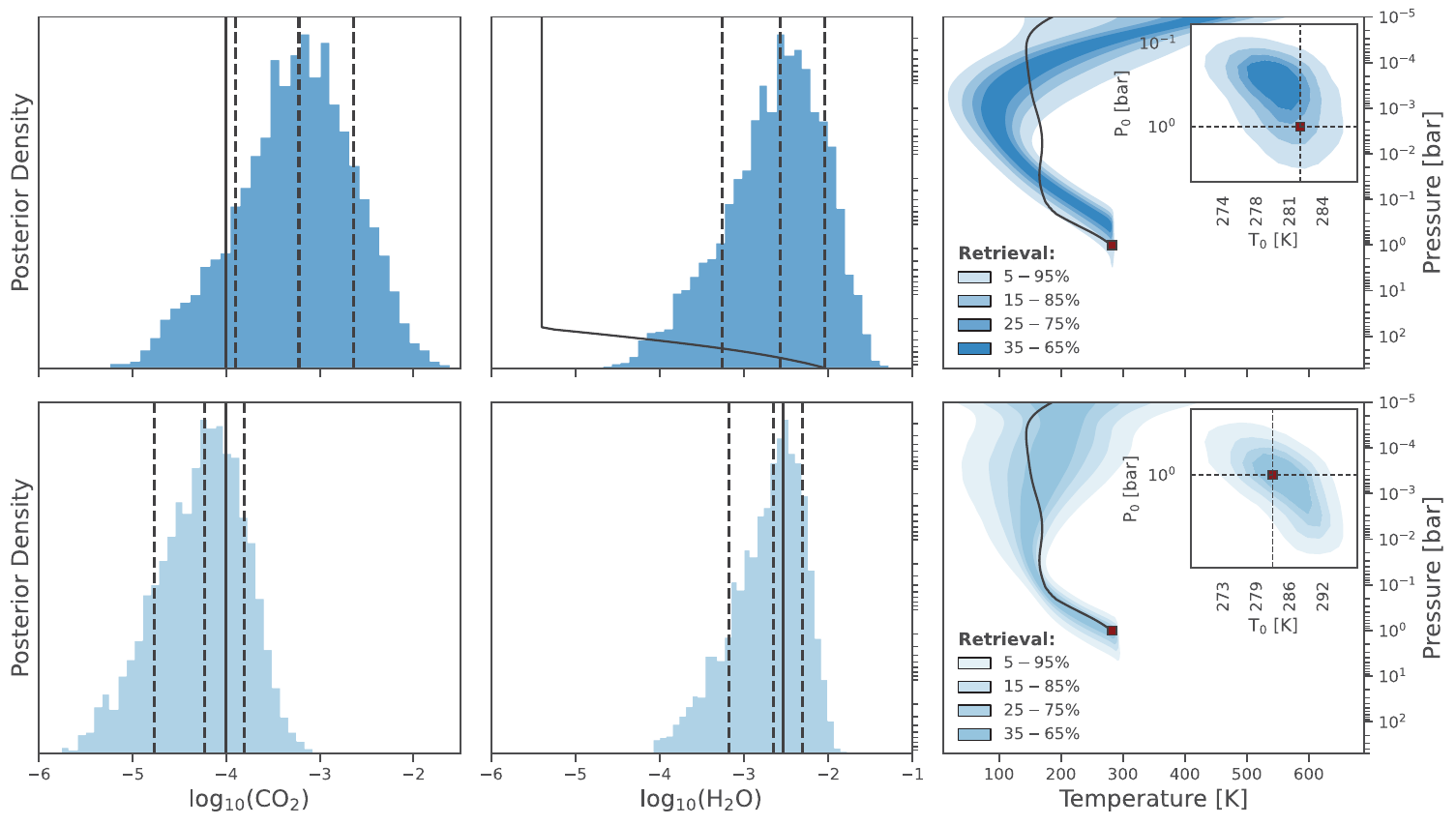}
    \caption{Comparison of atmospheric retrieval results between variable (top) and constant (bottom) $\mathrm{H_2O}$ profiles in the input spectrum of the p$\mathrm{CO_2}$\,=\,10$^{-4}$\,bar and $S/S_{\oplus}$\,=\,1.0 scenario, for an observational sensitivity of $S/N$\,=\,20 and $R$\,=\,100. We show, from left to right, retrieved $\mathrm{CO_2}$ abundance, $\mathrm{H_2O}$ abundance and $P-T$ profile constraints. Solid black lines indicate the ground truths, while dashed black lines denote the posterior median and 1$\sigma$ envelope. The inset in the $P-T$ profile panel illustrates the retrieved constraints on surface conditions.}
    \label{fig:H2O_var_fix}
\end{figure*}

\subsubsection{Treatment of \texorpdfstring{$H_2O$}{H2O} profiles in atmospheric retrievals}\label{subsec: H2O}
A contributing factor to systematic characterization biases, in form of offsets between retrieved posteriors and true p$\mathrm{CO_2}$, can be found in an incorrect parametrization of the $\mathrm{H_2O}$ abundance profile in the retrieval forward model. Here, we address the role of $\mathrm{H_2O}$ in constraining accurate $\mathrm{CO_2}$ abundances with atmospheric retrievals and how abundance constraints can be improved. To clarify, we first address the biases in abundance estimates, specifically focusing on the volume mixing ratio (VMR). Later, we examine how these biases affect the retrieved partial pressures.

$\mathrm{H_2O}$ has multiple strong absorption bands in the mid-infrared wavelength range and varies over orders of magnitude in its abundance with altitude (see Figure \ref{fig:spectra_species_profiles}). In our atmospheric population, we can additionally see a large variation in $\mathrm{H_2O}$ abundance profiles across our set of HZ characteristic atmospheres. It is, however, a common simplification of 1D atmospheric retrievals to assume vertically constant abundance profiles. While $\mathrm{CO_2}$ does not vary significantly with altitude, the assumption of pressure-independent abundance profiles is an especially strong simplification in the case of $\mathrm{H_2O}$.

To understand the impact of $\mathrm{H_2O}$ profile parametrization in the retrieval forward model on $\mathrm{CO_2}$ abundance constraints, we compare retrieval results from two scenarios. The first scenario involves input emission spectra with variable $\mathrm{H_2O}$ profiles, which corresponds to our original analysis. The second scenario assumes vertically constant $\mathrm{H_2O}$ abundances in the input atmospheres. We derive vertically constant abundances by calculating the pressure-weighted average of the variable $\mathrm{H_2O}$ abundance profiles. Figure \ref{fig:H2O_var_fix} illustrates the $\mathrm{CO_2}$ and $\mathrm{H_2O}$ abundance posteriors, along with the retrieved $P-T$ profiles for the p$\mathrm{CO_2}$\,=\,10$^{-4}$\,bar and $S/S_{\oplus}$\,=\,1.0 case. Here, we focus on the highest spectrum quality scenario.

When accounting for a variable $\mathrm{H_2O}$ profile in the input spectrum, we observe an offset in the $\mathrm{CO_2}$ posteriors toward larger than ground truth abundances, with $\mathrm{CO_2}$ overestimated by $\sim$1\,dex (Figure \ref{fig:H2O_var_fix}, top row). The retrieval struggles to accurately constrain $\mathrm{H_2O}$ abundance due to the strongly variable $\mathrm{H_2O}$ profile, yielding posteriors centered between surface and stratospheric abundances (see Appendix \ref{sec:app_supp_retrievals} for all scenarios). The retrieval selects an $\mathrm{H_2O}$ abundance and $P-T$ profile that best fit the strong $\mathrm{H_2O}$ features in the mid-infrared, leading to biased pressure estimates due to the incorrect treatment of the $\mathrm{H_2O}$ profile in the forward model. This results in an underestimated surface pressure and a shift of the retrieved $P-T$ profile toward lower than ground truth pressures (see Appendix \ref{sec:app_supp_retrievals} for all scenarios). Similar biases have been observed in previous studies using the same retrieval pipeline with pressure-dependent $\mathrm{H_2O}$ profiles in the input spectra \citep{alei2022, Mettler_2024}. The underestimation of pressure correlates with an overestimation of retrieved $\mathrm{CO_2}$ abundances due to a degeneracy between atmospheric species abundances and pressure broadening of spectral lines \citep{Misra_2014, Schwieterman_2015, alei2022, Mettler_2024}. This degeneracy between abundances and atmospheric pressure explains the overestimated $\mathrm{CO_2}$ content.
 
We then analyze results from input atmospheres with vertically constant $\mathrm{H_2O}$ abundances (Figure \ref{fig:H2O_var_fix}, bottom row). In these cases, $\mathrm{H_2O}$ abundances are accurately and precisely constrained, as our $\mathrm{H_2O}$ profile model matches the input profile. This improved $\mathrm{H_2O}$ estimation leads to retrieved thermal structures better centered around ground truths, which in turn results in $\mathrm{CO_2}$ abundance posteriors more accurately centered around true values. Retrieval results for the vertically constant $\mathrm{H_2O}$ input profiles are presented for the complete atmospheric population in Appendix \ref{sec:app_supp_retrievals}.

Further evidence supporting the impact of $\mathrm{H_2O}$ parametrization on pressure and $\mathrm{CO_2}$ constraint biases is provided by \citet{Konrad_2024}. The authors implemented a physical water condensation representation in the retrieval framework, resulting in improved characterization performance through a more accurate retrieval of atmospheric species abundances and thermal profiles for an Earth-twin case. However, preliminary attempts to constrain $\mathrm{H_2O}$ profiles with this refined parametrization for one of the atmospheric cases considered here showed limited characterization improvements. Future work could further test the flexibility of the physical water condensation framework on the generated atmospheric population and assess implications for detecting population-level $\mathrm{CO_2}$ trends. 

\begin{figure*}[tbp]
    \centering
    \includegraphics[width=\linewidth]{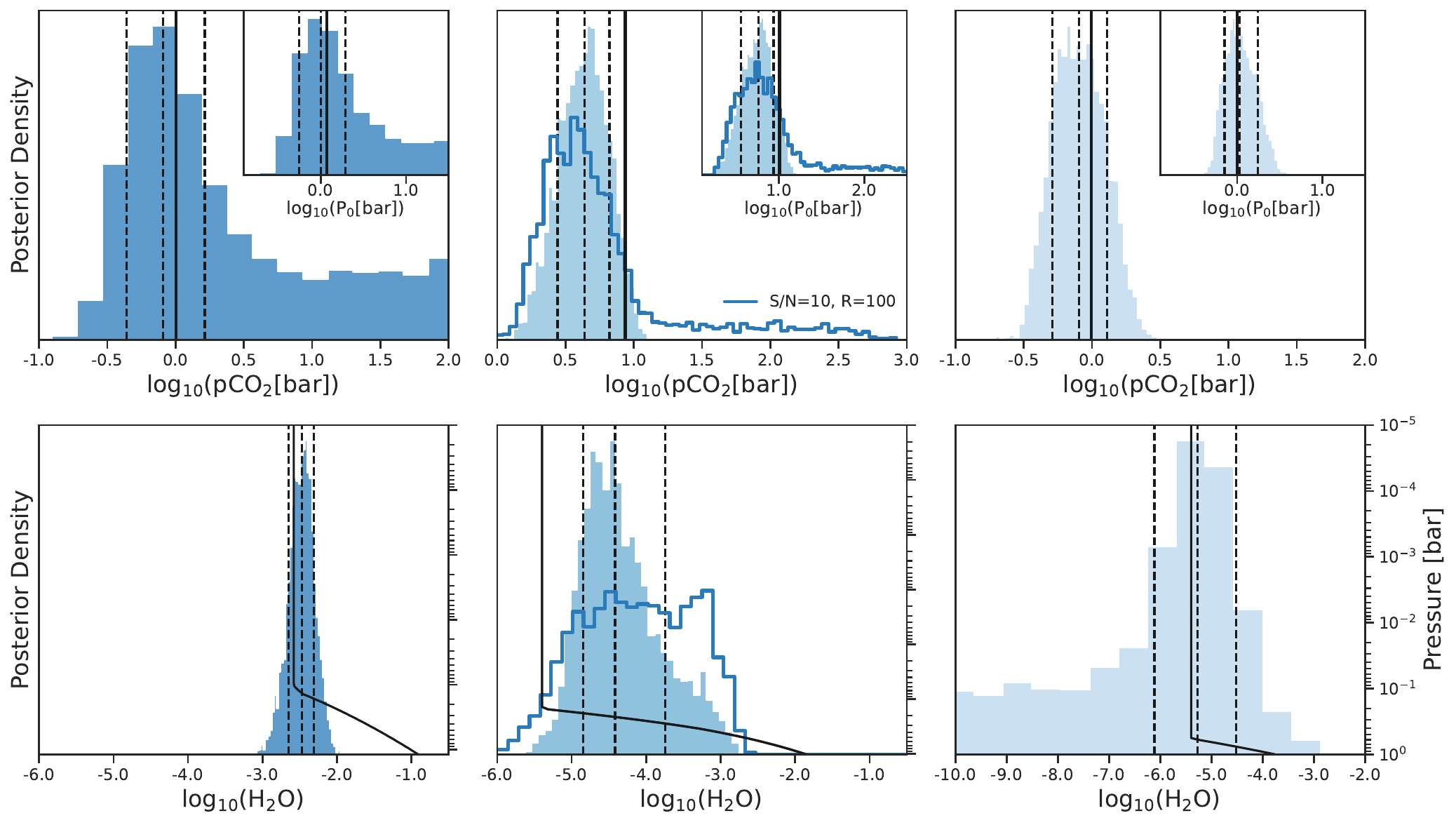}
    \caption{Atmospheric retrieval results for three $\mathrm{CO_2}$-dominated scenarios from our atmospheric population. Case 1 (left): p$\mathrm{CO_2}$\,=\,1\,bar, $S/S_{\oplus}$\,=\,1.0; Case 2 (center): p$\mathrm{CO_2}$\,=\,10\,bar, $S/S_{\oplus}$\,=\,0.4; and Case 3 (right): p$\mathrm{CO_2}$\,=\,10\,bar, $S/S_{\oplus}$\,=\,0.4. The top row illustrates the posterior distributions for the retrieved p$\mathrm{CO_2}$ and $P_0$ parameters, while the bottom row shows estimated $\mathrm{H_2O}$ abundances for each case. All retrievals were performed using simulated observations with $S/N$\,=\,20 and $R$\,=\,100. Solid black lines indicate the ground truths, while dashed black lines denote the posterior median and 1$\sigma$ envelope. In SL posterior cases, we show median and 1$\sigma$ envelope of the sensitivity peak.} 
    \label{fig:high_CO2}
\end{figure*}

\subsubsection{Characterization Potential of Atmospheres at the Habitable Limit}\label{subsec:hab_limit}
Constraining surface conditions in the p$\mathrm{CO_2}$\,$\geq$\,1\,bar parameter space introduces an additional characterization bias. We examine the characterization potential of $\mathrm{CO_2}$-dominated atmospheres at the habitable limit of our atmospheric population, focusing on retrieved constraints for $\mathrm{CO_2}$ partial pressures, surface pressures $P_0$, and $\mathrm{H_2O}$ abundances, as visualized in Figure \ref{fig:high_CO2}. Three habitable edge cases are analyzed - two at the high-temperature and one at the low-temperature limit.

\textbf{Case 1 (}$\bm{T_0}$\,$>$\,$\bm{330}$\,\textbf{K):} We derive SL posteriors for p$\mathrm{CO_2}$ and surface pressure $P_0$. While lower bounds for these parameters are retrieved, upper limits remain unconstrained. As outlined in Section \ref{subsec:results_postmaps}, p$\mathrm{CO_2}$ constraints are calculated by multiplying $P_0$ with $\mathrm{CO_2}$ abundance posteriors. Consequently, sensitivity limits on the surface pressure $P_0$ similarly influence SL-constraints on $\mathrm{CO_2}$ partial pressures. $\mathrm{H_2O}$ abundances are constrained to stratospheric levels, suggesting that only the upper atmosphere is being probed. This indicates that the high $\mathrm{H_2O}$ content ($\sim$\,10\%) and temperature (see Figure \ref{fig:spectra_species_profiles}) result in an optically thick lower atmosphere, restricting our ability to determine surface conditions. This limitation is consistent with the retrieved $P_0$ constraints.

\textbf{Case 2 (}$\bm{T_0}$\,$>$\,$\bm{330}$\,\textbf{K):} In this scenario, both $\mathrm{CO_2}$ partial pressures and surface pressure $P_0$ are underestimated, with the ground truth falling outside the 1$\sigma$ posterior envelope. This contrasts with observations in the p$\mathrm{CO_2}$\,$<$\,1\,bar parameter space, where underestimated $P_0$ constraints are associated with overestimated $\mathrm{CO_2}$ partial pressures. The p$\mathrm{CO_2}$ constraints correspond to the product of $P_0$ and $\mathrm{CO_2}$ VMR posteriors. In the p$\mathrm{CO_2}$\,$\geq$\,1\,bar parameter space, $\mathrm{CO_2}$ VMRs are tightly constrained at the upper prior edge, causing p$\mathrm{CO_2}$ posteriors to closely follow surface pressure constraints, explaining the tendency for negative offsets. The retrieved $\mathrm{H_2O}$ abundance constraint indicates atmospheric probing below stratospheric levels, corresponding to the non-constant regime of the $\mathrm{H_2O}$ abundance profile. This is typical for most atmospheric scenarios in our population. The incorrect treatment of the $\mathrm{H_2O}$ profile as vertically constant in the retrieval forward model leads to biased surface pressure constraints. Additionally, a spectrum quality dependence is observed for $\mathrm{CO_2}$ partial pressure and surface pressure constraints. At $S/N$\,=\,10, only SL constraints are obtained for both parameters, indicating reduced capabilities to infer surface conditions.  

\textbf{Case 3 (}$\bm{T_0}$\,$<$\,$\bm{250}$\,\textbf{K):} In this low temperature case, we are able to accurately constrain p$\mathrm{CO_2}$ (ground truth within 1$\sigma$ posterior envelope). Residing at the outer edge of the HZ with a $\mathrm{CO_2}$ partial pressure of 1\,bar, this atmosphere contains the lowest $\mathrm{H_2O}$ abundance within the atmospheric sample considered here. Driven by the low atmospheric $\mathrm{H_2O}$ content and low overall emitted flux of this cold outer HZ planet (see Figure \ref{fig:spectra_species_profiles}), we can only derive an upper limit for the $\mathrm{H_2O}$ abundance. In the context of constraining accurate $\mathrm{CO_2}$ partial pressures, low $\mathrm{H_2O}$ mixing ratios do not have further implications. 

In summary, negative p$\mathrm{CO_2}$ constraint offsets in the p$\mathrm{CO_2}$\,$\geq$\,1\,bar parameter space are similarly driven by the assumption of constant $\mathrm{H_2O}$ profiles within the retrieval. However, these offsets exhibit an opposite bias in p$\mathrm{CO_2}$ constraints compared to p$\mathrm{CO_2}$\,$<$\,1\,bar cases, attributable to the high $\mathrm{CO_2}$ content. $\mathrm{CO_2}$ abundances are tightly constrained at the upper prior limit, precluding overestimation. Hence, the retrieved p$\mathrm{CO_2}$ in this parameter space closely follows the $P_0$ constraint, which is typically underestimated due to the retrieval's difficulty in accurately determining $\mathrm{H_2O}$ abundances, as observed in most scenarios within our population. Furthermore, optically thick lower atmospheres naturally limit the ability to constrain surface conditions using mid-infrared thermal emission spectra. However, in the context of this study, the impact on p$\mathrm{CO_2}$ constraint capabilities remains limited, as we exclusively consider SL-limited posteriors for their Gaussian sensitivity peaks (see Section \ref{subsec:results_postmaps}). The $\mathrm{CO_2}$ partial pressures we retrieve here are sufficiently constrained to introduce only minor biases to our $\mathrm{CO_2}$ trend detectability assessment. 

\subsubsection{Implications for Trend Detection Capabilities}\label{subsec:trend_strat}
We identify characterization biases based on simplified species profile parametrization in retrieval forward models. Though the respective implications for $\mathrm{CO_2}$ partial pressure constraints vary depending on the atmosphere's location in the 2D p$\mathrm{CO_2}$-$S$ parameter space, these biases are visible across the atmospheric population investigated here. As a consequence, these biases translate to biased $\mathrm{CO_2}$ trend detections. We predict that atmospheric characterization biases might readily be reduced by applying a more accurate, physical parametrization of e.g. $\mathrm{H_2O}$, as an important opacity source in the mid-infrared spectral range that shows orders of magnitude variability within HZ characteristic atmospheres. This, necessarily, demands for forward models and retrieval routines in general, that are robust against population-scale atmospheric diversity.

In addition, the information content retrievable from thermal emission spectra is naturally limited by optically thick atmospheric layers. Especially at the high temperature edge of the HZ, we are likely to encounter large $\mathrm{H_2O}$ abundances driving optically thick lower atmospheres, masking deeper contributions to the spectrum. Although accurate constraints of high $\mathrm{CO_2}$ abundances might still be possible here, the inference of surface conditions is rather challenging in these cases. A potential lack of well-constrained surface conditions at the high temperature habitability edge may reduce the parameter space of survey populations accessible to trend detection assessments.

\subsection{Limitations}\label{subsec:limitations}
In this study we show that population-level $\mathrm{CO_2}$ trends can readily be detected with LIFE observations of a population of terrestrial HZ atmospheres. In contrast, differentiating biotic from abiotic $\mathrm{CO_2}$ trends poses challenges that require advancements in our present atmospheric characterization frameworks. While LIFE serves as a specific case study, these findings are widely relevant to any future space-based mid-infrared mission designed to perform these population-scale atmospheric characterizations. To set these results into broader context, we outline the limitations of this work.

First, our atmospheric model assumes a simplified composition consisting of $\mathrm{CO_2}$, $\mathrm{H_2O}$ and $\mathrm{N_2}$. However, we can expect actual atmospheric populations to show greater diversity in compositions than considered here (for a review, see, e.g., \citealt{Wordsworth_2022}). The inclusion of additional species, such as $\mathrm{CH_4}$ or $\mathrm{O_3}$, would influence the self-consistent modeling of planetary atmospheres, impacting thermal structures and surface conditions. Moreover, the presence of other atmospheric components may affect the ability of the retrieval to constrain $\mathrm{CO_2}$ abundances in multiple ways. For instance, overlapping absorption bands, such as those of $\mathrm{O_3}$ and $\mathrm{CO_2}$ (both at $\sim$\,10\,$\mu$m) or $\mathrm{CH_4}$ (around 8\,$\mu$m) and $\mathrm{H_2O}$ (ranging from 5 to 8\,$\mu$m) (e.g., \citealt{konrad2022}), along with the increased complexity introduced by additional parameters in the retrieval, could impact the precision of $\mathrm{CO_2}$ abundance estimates.

The influence of clouds is neglected in both the input spectra and the retrieval forward model. This assumption is particularly relevant in the high surface temperature regime, where clouds would impact atmospheric characterization due to the expected high atmospheric water content. However, clouds have a significantly smaller effect on mid-infrared radiation than on reflected stellar light (e.g. \citealt{Kitzmann_2011, Marley_2013, Fauchez_2019}). Nevertheless, previous studies have examined the performance of the retrieval framework employed in this study in the context of cloudy Earth-like emission spectra. On Earth, patchy $\mathrm{H_2O}$ cloud coverage partially obstructs thermal emissions from lower atmospheric layers. Both \citet{alei2022} and \citet{Mettler_2024} conducted atmospheric retrievals on cloudy emission spectra while assuming a cloud-free atmosphere, leading to biases in their inferred atmospheric and planetary parameters. Specifically, \citet{Mettler_2024} identified biases in the retrieved surface temperature and planetary radius estimates, which were attributed to the omission of cloud effects. These findings were further validated by \citet{Konrad_2024}, who demonstrated that incorporating clouds into the retrieval forward model mitigates biases in the retrieved temperature and radius estimates. In addition to $\mathrm{H_2O}$ cloud formation, the condensation of $\mathrm{CO_2}$ and the subsequent development of $\mathrm{CO_2}$ clouds become significant at the outer edge of the HZ, leading to a complex interplay between radiative warming and cooling effects on planetary climates (e.g., \citealt{Forget_1997, Kitzmann_2017}). On this basis, future studies should explore the implications of both $\mathrm{H_2O}$ and $\mathrm{CO_2}$ clouds for the performance of atmospheric retrievals applied across a broad population of HZ characteristic atmospheres and investigate their impact on the detectability of population-level $\mathrm{CO_2}$ trends.

Moreover, we neglect photochemical processes in our HZ characteristic atmospheric population. In Earth's atmosphere we observe a considerable altitude-dependence of certain species, e.g. $\mathrm{H_2O}$ and $\mathrm{O_3}$, which are driven by photochemical reactions in the stratosphere (e.g. \citealt{Chapman_1932, Jones_1986}). Consequently, assuming constant abundance profiles may introduce biases in the retrieved abundance estimates. In contrast, $\mathrm{CO_2}$ is well mixed throughout the atmosphere, making the assumption of a constant abundance profile a sufficient approximation (e.g., \citealt{Rugheimer_2018, Mettler_2024}). Omitting photochemistry may further impact the self-consistent modeling of planetary atmospheres. For instance, in the case of Earth, photochemical $\mathrm{O_3}$ induces a temperature inversion in the $P-T$ profile, which subsequently influences Earth's mid-infrared emission spectrum. 

In our retrieval pipeline we parametrize atmospheric thermal structures with a fourth order polynomial. This model allow us to be flexible in the types of $P-T$ profiles that we fit within the retrieval. However, the higher the number of parameters is, the more computationally expensive the retrieval becomes. In context of providing an accurate thermal structure fit with a relatively low number of parameters, a learning-based $P-T$ profile parametrization could provide a powerful alternative \citep{Gebhard_2024}. Additionally, the learning-based approach would guarantee physical thermal profiles, whereas a polynomial parametrization cannot inherently exclude unphysical ones. Currently, there is, however, a lack of $P-T$ profile training data in the temperate terrestrial atmosphere regime which limits the reliability of this approach. Especially in context of atmospheric characterization on a population-level, these models need to be equipped with a broad swath of thermal structures to depict a wide atmospheric variety.

Our grid of 12 atmospheric states is a discrete representation of characteristic HZ atmospheres and cannot represent the full atmospheric diversity that can be expected in real terrestrial HZ atmospheres. We assume all planets to be located at 10\,pc from the observer orbiting a G2V star and consisting of Earth-like planetary parameters, such as planetary radius and surface gravity. This approach is justified, since the grid has the primary function to interpolate retrieval results without the need to perform large-scale, computationally intensive, retrievals on the order of population sizes ($N_P$\,$\leq$\,100) that we investigate here. Naturally, the estimated p$\mathrm{CO_2}$ posterior maps cannot be translated to arbitrary atmospheric cases but need to be understood within the context of our atmospheric population and respective assumptions. Furthermore, the accuracy of interpolating between posteriors of various atmospheric scenarios as a proxy for direct retrieval results remains uncertain, warranting further investigation to understand the potential and limitations of this approach.

The aim of the present study is to test observational capabilities to detect and constrain population-level $\mathrm{CO_2}$ trends. However, \citet{Lehmer2020} suggested differentiating between the 2D p$\mathrm{CO_2}$-$S$ population and a uniform population within habitable temperature limits to achieve an unambiguous test for a functioning Cb-Si cycle. A linear null trend, following the null (uniform) population defined after \citet{Lehmer2020}, would be naturally more challenging to reject than a flat trend. However, this null trend definition relies on assumptions of shape and extent of a null population. Given the lack of a well-characterized Cb-Si null population, we consider the absence of correlation (i.e., a flat trend) in $\mathrm{CO_2}$ partial pressures with stellar irradiation to be the more robust null hypothesis in our trend assessment context.

Besides detecting population-level $\mathrm{CO_2}$ trends, we assess the potential to differentiate biotic from abiotic $\mathrm{CO_2}$ trends. Given that we have only one instance of a Cb-Si world that, coincidentally or not, evolved to host a biosphere, we lack predictions for a corresponding abiotic scenario. Therefore, as previously outlined, we restrict the biotic enhancement of chemical weathering in the Cb-Si model \citep{Kriss-Tott2017, Kriss-Tott2018, Lehmer2020} to generate a second p$\mathrm{CO_2}$-$S$ population with minimal biotic influence on the Cb-Si weathering feedback mechanism. Consequently, our biotic Cb-Si distribution encompasses a broad variability of biotic impact on the weathering feedback, with the abiotic distribution corresponding to a subset of this. However, these distributions likely do not provide an accurate representation of reality. We might observe a stronger mix of Cb-Si populations with decisive and weak biotic impact, a mix of worlds with and without functioning Cb-Si cycles or potentially no indicators of a trend in $\mathrm{CO_2}$ abundance with stellar irradiation whatsoever. Future work could investigate biases in trend detection by exploring more complex mixed populations, which would still require assumptions about the respective biotic, abiotic, and null Cb-Si distributions. Importantly, the Cb-Si populations considered here serve as cases to test observational capabilities against. Thereby, we initiate a pathway toward gaining insight into the actual presence and characteristics of Cb-Si driven $\mathrm{CO_2}$ trends by equipping future missions with the necessary capabilities to ultimately test these hypotheses empirically. 

\section{Summary and Conclusions}\label{sec:conclusions}
In this study, we investigated how a trend in $\mathrm{CO_2}$ abundance with stellar irradiation can be detected from thermal emission spectra of a population of temperate terrestrial exoplanet atmospheres. We explored this trend as tracer for the Cb-Si weathering cycle, a population-level habitability signature. Additionally, we assessed the ability to differentiate between $\mathrm{CO_2}$ trends in atmospheric populations with global-scale biospheres and largely abiotic worlds, examining the observational feasibility of $\mathrm{CO_2}$ trends as population-level indicators of global biotic activity. We created synthetic exoplanet survey populations with atmospheric compositions based on geochemistry-climate predictions and performed retrievals on simulated LIFE observations of the planets' thermal emission. Using a hierarchical importance sampling formalism, we inferred trend constraints, assessing the diagnostic power of a population-scale characterization approach.

We observe a robust detection of $\mathrm{CO_2}$ trends for population sizes $N_P$\,$\geq$\,30 and all considered spectrum quality scenarios $S/N$\,=\,[10,\,20] and $R$\,=\,[50,\,100] in both biotic and abiotic cases. The strength of trend detection is predominantly influenced by $N_P$, while largely independent of the considered $S/N$ and $R$ assumptions. We obtain overall higher confidence in trend detection in the biotic than the abiotic case. This observation is associated with the difference in p$\mathrm{CO_2}$-$S$ parameter space covered by predicted biotic and abiotic Cb-Si distributions. Notably, we observe a tendency to underestimate $\mathrm{CO_2}$ trend slopes, which does not play a major limiting factor in trend detection but influences the assessed capabilities to correctly differentiate biotic from abiotic $\mathrm{CO_2}$ trends.

In context of differentiating between biotic and abiotic trends, we find that both accuracy (centering of trend slope estimate around true trend) and precision (tightness of trend estimate) play significant roles. We find the retrieved trend slope constraints to be consistently offset from the ground truths, which leads to a biased characterization of biotic and abiotic $\mathrm{CO_2}$ trends. By investigating drivers of these trend offsets, we find biases in individual planet p$\mathrm{CO_2}$ constraints to be a major factor for trend constraint accuracy, whereas the number of characterized atmospheres $N_P$ as well as the underlying variability of true p$\mathrm{CO_2}$ within the population represent significant factors for trend constraint precision. If p$\mathrm{CO_2}$ constraint biases were corrected, biotic and abiotic population trends could be accurately constrained for $N_P$\,$\geq$\,100.

We identify two classes of p$\mathrm{CO_2}$ constraint biases. The first bias is driven by the widely applied simplification of vertically constant atmospheric species profiles in the retrieval forward models. Here, $\mathrm{H_2O}$ plays a significant role due to its large variability in HZ characteristic atmospheres and its strong mid-infrared opacity which greatly influences the retrieval performance and respective constraints on $\mathrm{CO_2}$ abundances. However, further advancements are needed to develop retrieval frameworks that remain robust against population-scale atmospheric diversity. The second p$\mathrm{CO_2}$ constraint bias is driven by limited capabilities to retrieve surface pressure constraints on high surface temperature habitability edge cases. Here, large water vapor abundances lead to optically thick lower atmospheres, which naturally prohibit the probing of surface layers via thermal emission spectra. While this bias has only limited impact on our trend assessment here, it represents an inherent characterization limitation for atmospheres at the upper-temperature bound of the HZ.

We demonstrate the ability of future missions like LIFE, or similar mid-infrared interferometer concepts, to enable population-level characterization of temperate terrestrial atmospheres and find that Cb-Si cycle driven $\mathrm{CO_2}$ trends, as population-wide habitability signature, can readily be detected in a modest population of thermal emission spectra. However, we are only beginning to explore the potential of comparative planetology in terrestrial exoplanet atmospheres. Further studies, which test atmospheric characterization performance against broad atmospheric diversity, are essential to prepare next-generation observational facilities to provide robust and accurate constraints of atmospheric as well as planetary parameters. Then, efforts like these will pave the way toward assessing the commonness of habitable worlds or even global-scale biospheres outside of our Solar System.

\newpage
J.H., S.P.Q., and D.V. thank the ETH Centre for Origin and Prevalence of Life for its generous financial support. Parts of this work have been carried out within the framework of the National Centre of Competence in Research PlanetS supported by the Swiss National Science Foundation under grants 51NF40\_205606, 51NF40\_182901, and 200020\_200399. J.H., S.P.Q, D.A., B.S.K., E.G, and F.A.D. acknowledge the financial support of the SNSF. E.A.’s research was supported by an appointment to the NASA Postdoctoral Program at the NASA Goddard Space Flight Center, administered by Oak Ridge Associated Universities under contract with NASA. The authors would also like to thank Owen Lehmer, Joshua Krissansen-Totton, Jacob Lustig-Yaeger and Sandra Bastelberger for their expertise and useful discussions.

\textit{Author Contributions:}  J.H. carried out the analyses, created the figures, developed the software, and wrote the original manuscript. S.P.Q., D.A., and D.V. initiated the project and provided regular guidance, supervision, and conceptual input. B.S.K. and E.G. provided expertise on atmospheric retrieval and statistical analysis, respectively. J.K. contributed synthetic planet population data generated with \texttt{P-Pop}. All authors reviewed and edited the manuscript.

\textit{Data Availability:} All source data underlying the results of this study, including spectral data and retrieved parameter posterior distributions, have been archived on Zenodo \citep{Hansen_2025_Data}. The dataset is accessible at \href{https://zenodo.org/records/15056424}{https://zenodo.org/records/15056424}.

\textit{Software:} This work has made use of several open source packages, including \texttt{matplotlib} \citep{Hunter_2007}, \texttt{numpy} \citep{Harris_2020}, \texttt{pandas} \citep{Mckinney_2010}, petit\texttt{RADTRANS} \citep{molliere2019, Molliere_2020, alei2022}, and \texttt{scipy \citep{Virtanen_2020}}.


\clearpage
\appendix

\section{Trend Statistics: Preliminary Hypothesis Testing}\label{sec:hypothesis_testing}
To obtain a first understanding of the observational feasibility associated with detecting and differentiating biotic and abiotic $\mathrm{CO_{2}}$ trends in a population of temperate terrestrial atmospheres, we perform a set of preliminary frequentist hypothesis tests which assume a fixed population-wide observational constraint on $\mathrm{CO_{2}}$ partial pressures. This approach allows for a systematic evaluation of the relationship between observational uncertainty, sample size, and the ability to detect atmospheric $\mathrm{CO_{2}}$ trends across diverse planetary populations. 

\subsection{Hypothesis Test Setup}
We infer the required number of planets to detect atmospheric $\mathrm{CO_{2}}$ trends based on the assumed observational uncertainty of atmospheric $\mathrm{CO_{2}}$ partial pressure. Synthetic survey populations with up to $N_P$ = 100 planets per sample are constructed. For each $N_P$ scenario, incident flux values $S$ are drawn for each planet from a characteristic EEC distribution estimated from simulated LIFE observations (Section \ref{subsec:flux_dist}, Figure \ref{fig:flux_dist}). Each planet is then assigned a true atmospheric p$\mathrm{CO_{2}}$ value, dependent on its received stellar irradiation. The p$\mathrm{CO_{2}}$ level is drawn from either biotic or abiotic 2D p$\mathrm{CO_{2}}$-$S$ distributions (Section \ref{subsec:co2_trend}, Figure \ref{fig:bio_abio_dist}), depending on the scenario being tested. Each (p$\mathrm{CO_{2}}$, $S$) pair is associated with a simplified observational uncertainty $\delta$p$\mathrm{CO_{2}}$\,=\,±[0.1, 0.5, 1.0, 1.5] dex around the true p$\mathrm{CO_{2}}$, with the uncertainty level fixed across a given survey population. For each population size and observational uncertainty scenario, (p$\mathrm{CO_{2}}$, $S$) pairs are redrawn 10,000 times using bootstrapping with sample replacement. A weighted least squares (WLS) fit is performed for each instance to enable marginalization over the trend population for given sample size and uncertainty.
In order to infer capabilities to detect and differentiate biotic and abiotic $\mathrm{CO_{2}}$ trends, we conduct t-tests on the slope parameter, based on the following three hypothesis:

\begin{itemize}
    \item $\mathcal{H}_0$\,:\,$\beta$\,=\,0, $\mathcal{H}_1$\,:\,$\beta$\,$\neq$\,0: We test the detectability of a semi-logarithmic $\mathrm{CO_{2}}$ trend by testing the rejection of a null (flat) trend, where $\beta$ represents the logarithmic slope. This test is performed for both biotic and abiotic scenarios.
    \item $\mathcal{H}_{0,bio}$\,:\,$\beta$\,=\,$\beta_{abio}$, $\mathcal{H}_{1,bio}$\,:\,$\beta$\,$\neq$\,$\beta_{abio}$: We assess the differentiation between biotic and abiotic $\mathrm{CO_{2}}$ trends by testing whether an abiotic trend can be rejected in the biotic scenario.
    \item $\mathcal{H}_{0,abio}$\,:\,$\beta$\,=\,$\beta_{bio}$, $\mathcal{H}_{1,abio}$\,:\,$\beta$\,$\neq$\,$\beta_{bio}$: We assess the differentiation between biotic and abiotic $\mathrm{CO_{2}}$ trends by testing whether a biotic trend can be rejected in the abiotic scenario.
\end{itemize}

A hypothesis is considered rejected if the estimated p-value\,$<$\,0.05. We conduct two-tailed tests to allow for detection sensitivity of either effect direction. These tests are conducted on 10,000 fitted trends per survey population size and observational uncertainty case, yielding a set of 10,000 test outcomes for each $N_P$ and $\delta$p$\mathrm{CO_{2}}$ scenario. We estimate the minimum number of planets required to achieve a given fraction (statistical power) of the 10,000 tests that correctly reject the respective false null hypotheses (p-value\,$<$\,0.05). This approach allows us to examine the necessary population size and observational uncertainty levels required to achieve 95\%, 99\%, and 99.9\% likelihood of identifying trends, assuming these trends are present within the atmospheric population (see \citet{checlair_thesis_2021} for a comparable trend statistical analysis framework). 

The left panel in Figure \ref{fig:pvals_comb} illustrates the inference of minimum population sizes based on our hypothesis testing. This panel displays test outcomes for rejecting a flat trend (first hypothesis test) in the biotic scenario. The intersection of curves representing equal statistical power and the statistical significance threshold indicates the required magnitude of $N_P$. For instance, to achieve a 95\% rate of correctly rejecting the false null hypothesis of a flat trend, assuming $\delta$p$\mathrm{CO_{2}}$\,=\,1.0\,dex, a minimum of 23 EEC observations is necessary.

\subsection{Hypothesis Test Results: Sensitivity of \texorpdfstring{$\mathrm{CO_2}$}{CO2} Trend Detection and Differentiation}
\begin{figure*}[htb]
    \centering
    \includegraphics[width=\linewidth]{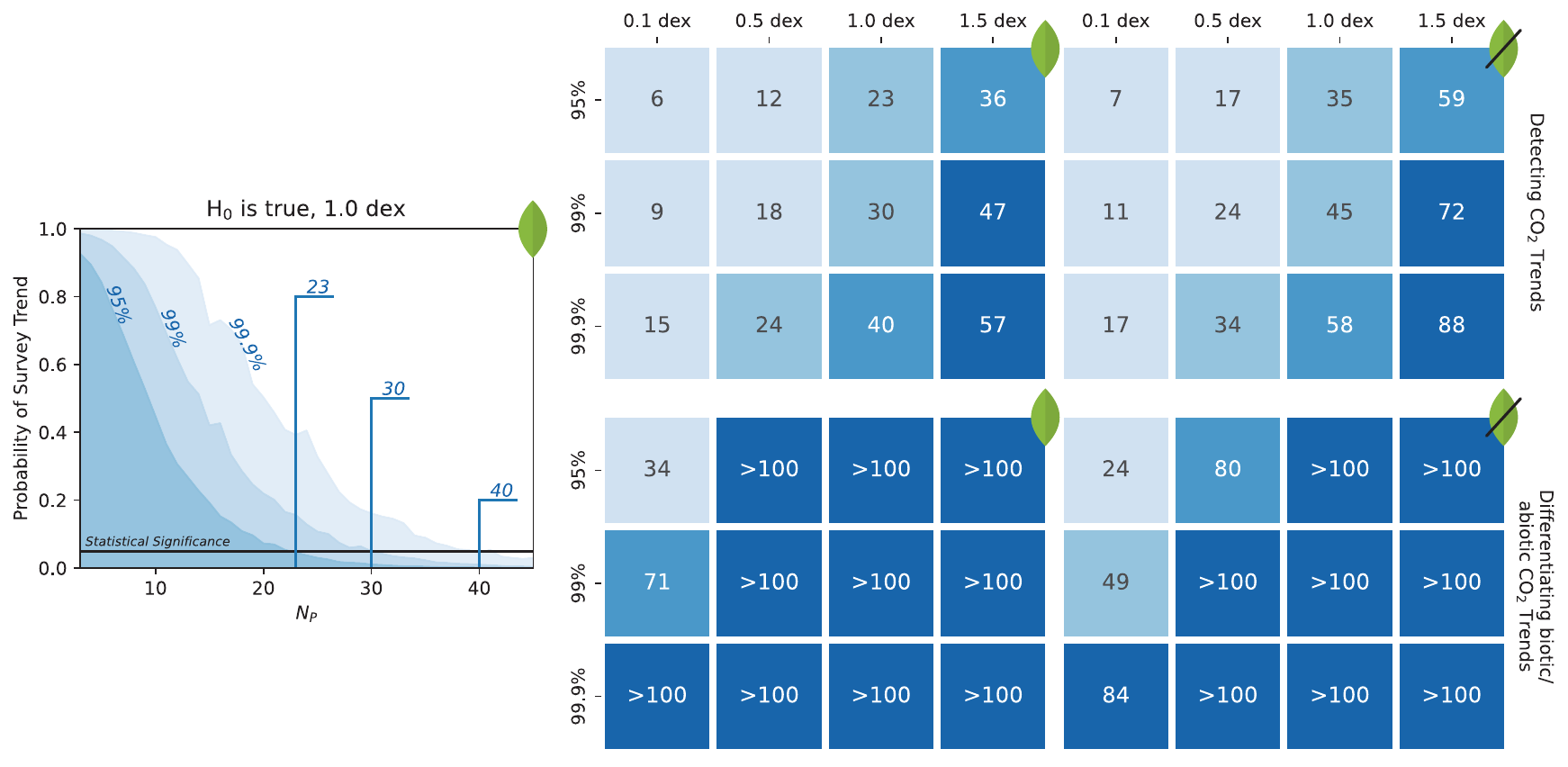}
    \caption{Minimum population sizes required to detect and differentiate biotic and abiotic $\mathrm{CO_{2}}$ trends. Left: Visualization of the minimum population size inference approach. Right: Map of minimum required population sizes $N_P$ to achieve statistical power between 95\% and 99.9\% across various observational uncertainty scenarios $\delta$p$\mathrm{CO_{2}}$. Upper panels represent trend detection (first hypothesis test), while lower panels illustrate differentiation between biotic and abiotic trends (second and third hypothesis tests). Color gradients indicate the magnitude of $N_P$ required under different $\delta$p$\mathrm{CO_{2}}$ and statistical power scenarios.}
    \label{fig:pvals_comb}
\end{figure*}

First we investigate the detectability of population-level $\mathrm{CO_{2}}$ trends in both biotic and abiotic scenarios, focusing on the minimum required population sizes $N_P$ to achieve a statistical power between 95\% and 99.9\% across various observational uncertainty scenarios. The analysis shown in the top row of Figure \ref{fig:pvals_comb} illustrates results for p$\mathrm{CO_{2}}$ drawn from biotic (left) and abiotic (right) Cb-Si distributions. Key findings indicate that larger observational uncertainties in atmospheric $\mathrm{CO_{2}}$ partial pressure necessitate higher numbers of planets to reject a null trend at a given statistical power. Additionally, fewer planets are required to reject the null trend in biotic distributions compared to abiotic distributions, attributed to the flatter shape of the abiotic distribution in the 2D p$\mathrm{CO_{2}}$-$S$ parameter space (Figure \ref{fig:bio_abio_dist}). For an observational uncertainty of $\delta$p$\mathrm{CO_{2}}$\,=\,0.5\,dex, $\geq$\,34 planets are required to reject a flat population-level $\mathrm{CO_{2}}$ trend at the highest considered statistical power in both biotic and abiotic cases.

We also assess the ability to differentiate between biotic and abiotic $\mathrm{CO_{2}}$ trends (Figure \ref{fig:pvals_comb}, bottom row), revealing larger required population sizes to reject respective null hypotheses compared to rejecting a null trend. Fewer planets are needed to reject a biotic trend in the abiotic scenario than vice versa, due to the distribution and location of the respective 2D p$\mathrm{CO_{2}}$-$S$ distributions. Since the abiotic distribution corresponds to a subpopulation of the biotic one, an abiotic trend is harder to reject when sampling from the (full) biotic parameter space. For observational uncertainties $\geq$\,0.5\,dex, both null hypothesis rejections generally require $>$\,100 observed EEC atmospheres.

In conclusion, detecting population-level $\mathrm{CO_{2}}$ trends requires fewer observations or allows for larger observational uncertainties compared to differentiating between biotic and abiotic population trends. Our hypothesis tests highlight a strong sensitivity of trend detectability to both observational uncertainty and available planet population sizes, as well as differences in trend detection capabilities between biotic and abiotic cases, which are linked to the respective distributions in the 2D p$\mathrm{CO_{2}}$-$S$ parameter space. These observed sensitivities underscore the importance of more nuanced estimates of observational uncertainty, as demonstrated in our main trend detection analysis, and emphasize the necessity of assessing broad survey population sizes to effectively address both trend detection and differentiation between biotic and abiotic trends.

\clearpage
\section{Supplementary Atmospheric Retrieval Results}\label{sec:app_supp_retrievals}
\begin{figure*}[!h]
    \centering
    \includegraphics[width=\linewidth]{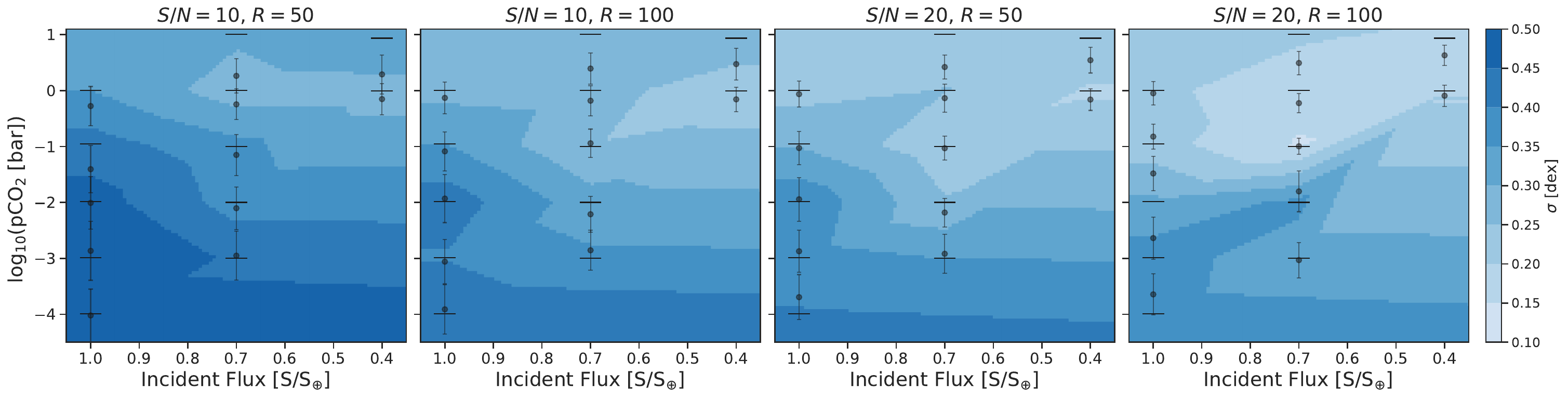}
    \caption{As top panel in Figure \ref{fig:sigma_delta_map} but for spectrum quality cases $S/N$ = [10,\,20] and $R$ = [50,\,100].} 
    \label{fig:sigma_map_all}
\end{figure*}

\begin{figure*}[!h]
    \centering
    \includegraphics[width=\linewidth]{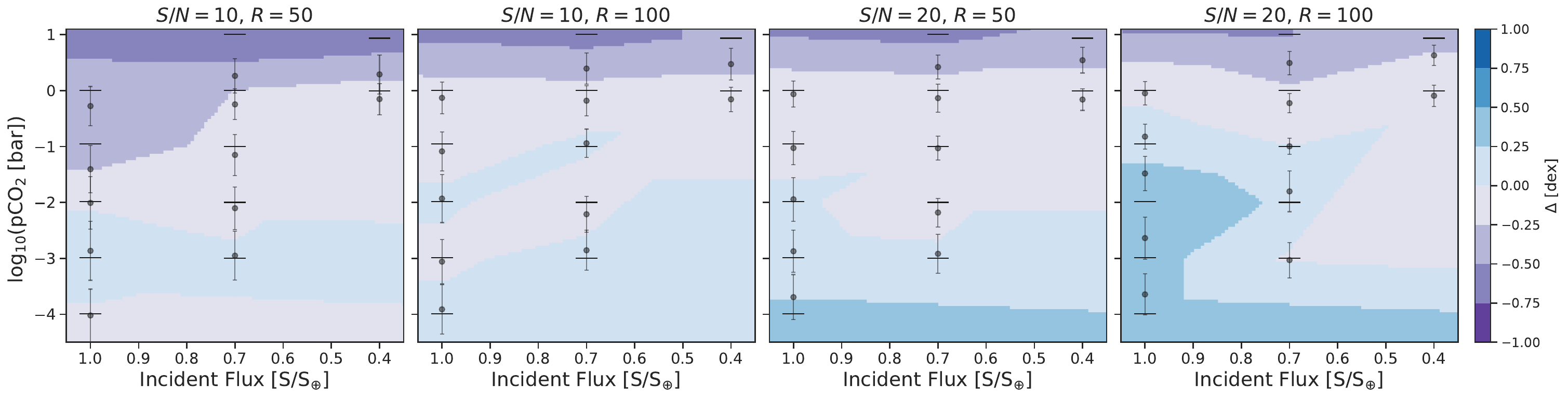}
    \caption{As bottom panel in Figure \ref{fig:sigma_delta_map} but for spectrum quality cases $S/N$ = [10,\,20] and $R$ = [50,\,100].} 
    \label{fig:delta_map_all}
\end{figure*}

\begin{figure*}[!h]
    \centering
    \includegraphics[width=0.9\linewidth]{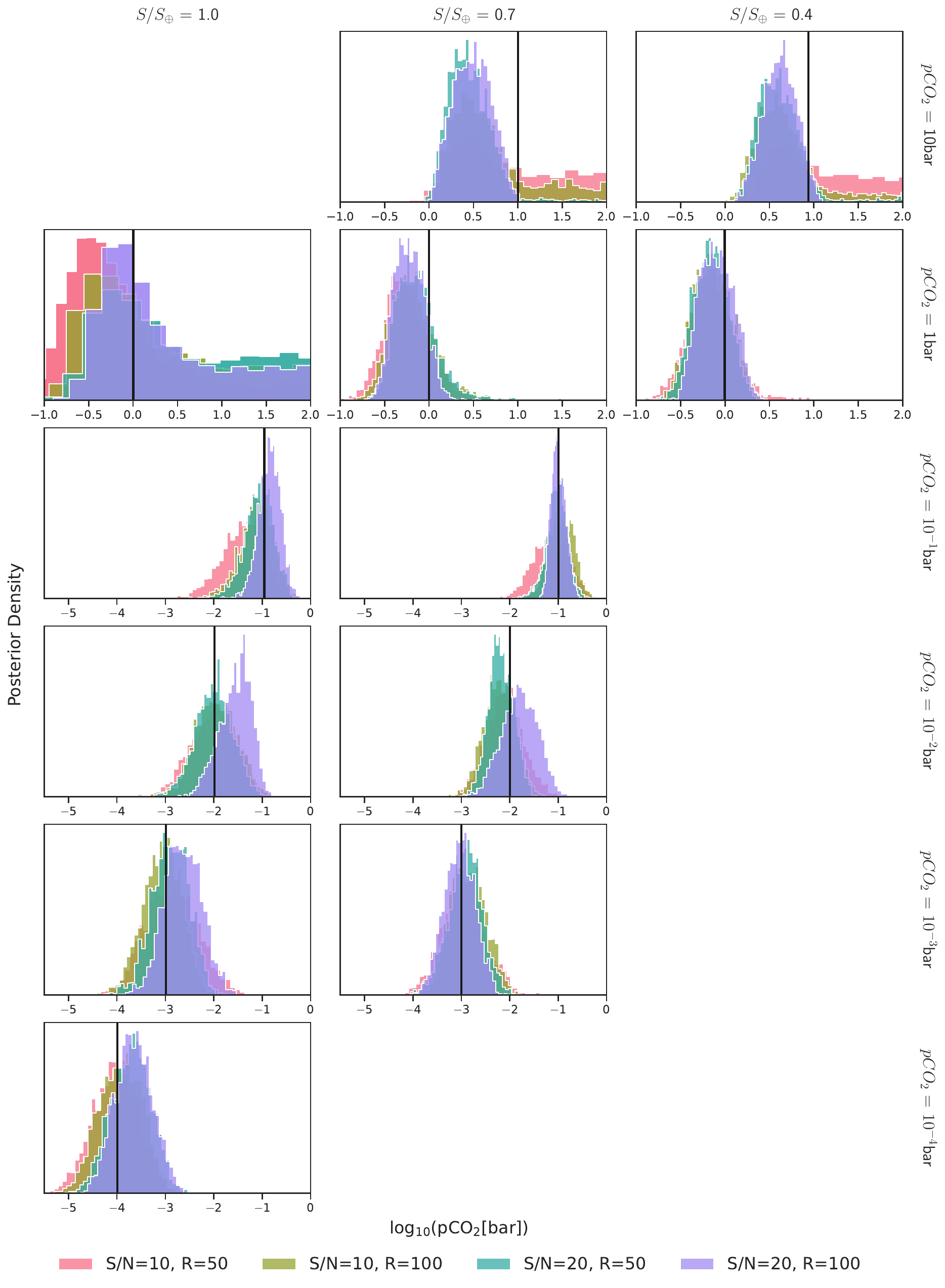}
    \caption{As Figure \ref{fig:CO2_post} but for $\mathrm{CO_2}$ partial pressures.}
    \label{fig:CO2_post_p}
\end{figure*}

\begin{figure*}[!h]
    \centering
    \includegraphics[width=0.84\linewidth]{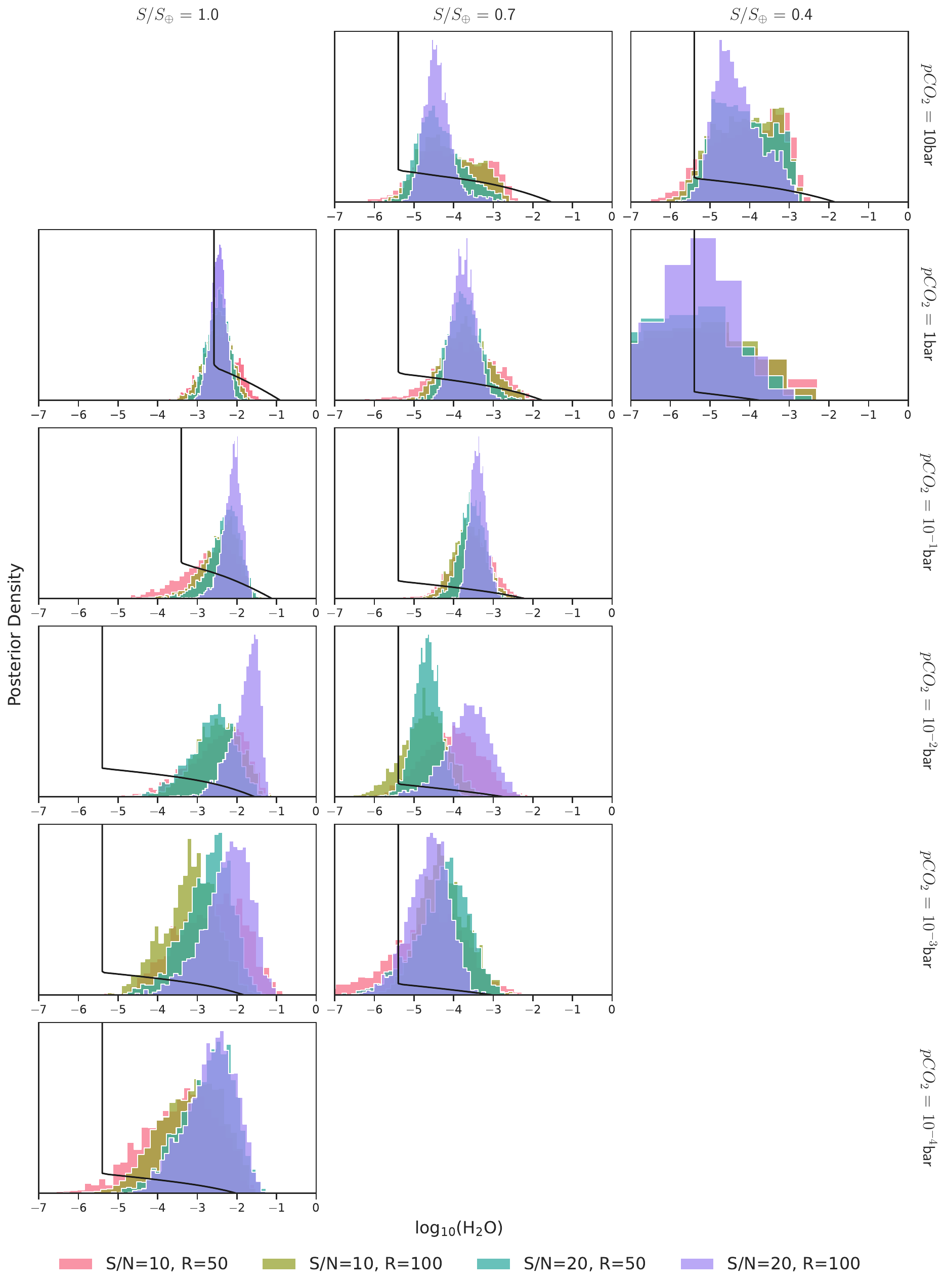}
    \caption{Posterior distributions of $\mathrm{H_2O}$ abundance derived from atmospheric spectrum retrievals. The panels are arranged to reflect the atmospheric grid within the p$\mathrm{CO_2}$-$S$ parameter space. Columns represent different stellar insolation values, $S/S_{\oplus}$\,=\,1.0, 0.7, 0.4, while rows correspond to different $\mathrm{CO_2}$ partial pressures, p$\mathrm{CO_2}$\,=\,10\,bar, 1\,bar, $10^{-1}$\,bar, $10^{-2}$\,bar, $10^{-3}$\,bar, $10^{-4}$\,bar. Each panel shows results for four combinations of signal-to-noise ratio, $S/N$\,=\,[10,\,20], and spectral resolution, $R$\,=\,[50,\,100]. Thick black lines indicate the ground truth $\mathrm{H_2O}$ abundance.} 
    \label{fig:H2O_post}
\end{figure*}

\begin{figure*}[!h]
    \centering
    \includegraphics[width=0.79\linewidth]{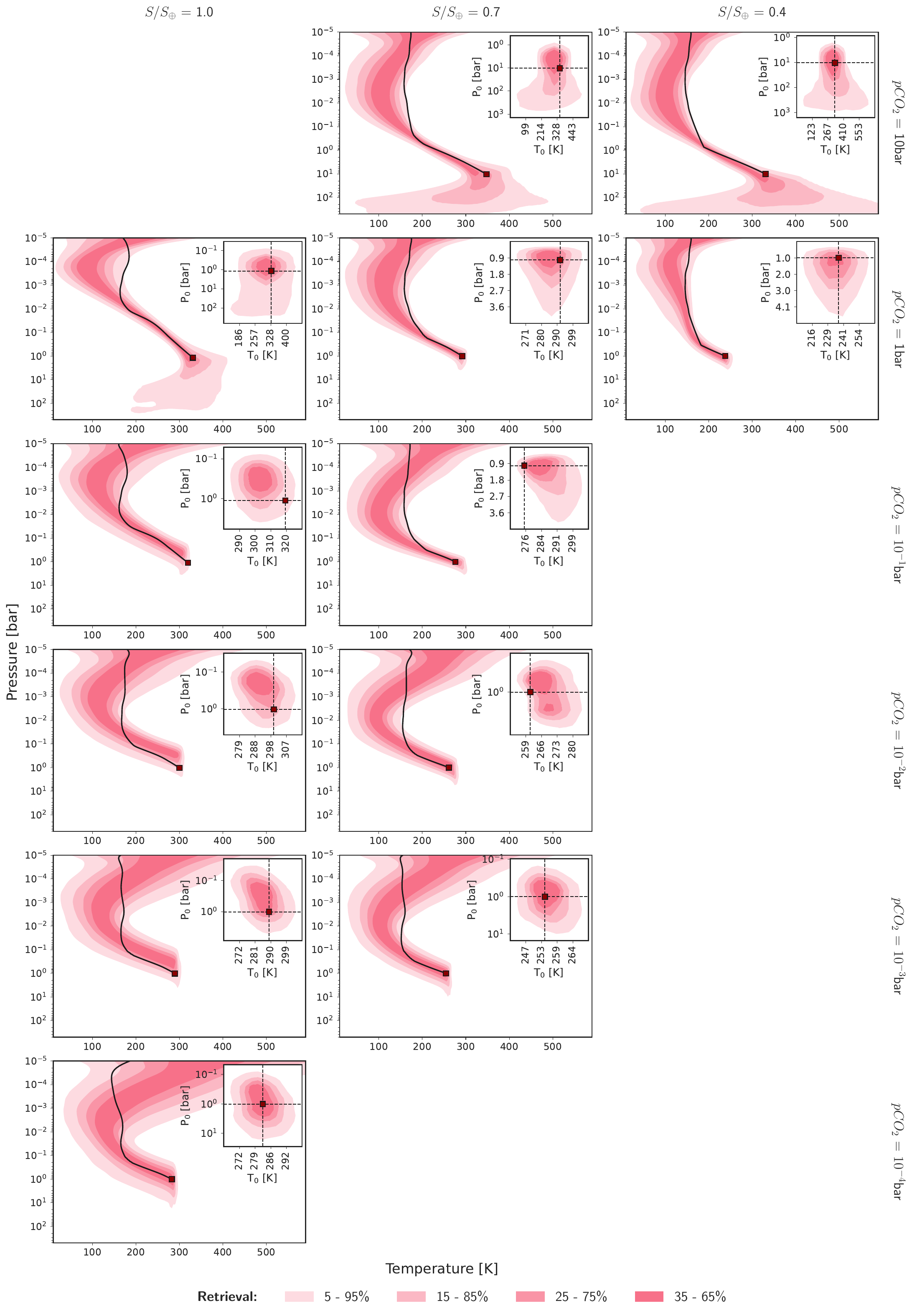}
    \caption{$P-T$ profiles derived from atmospheric spectrum retrievals for $S/N$\,=\,10 and $R$\,=\,50. The panels are arranged to reflect the atmospheric grid within the p$\mathrm{CO_2}$-$S$ parameter space. Columns represent different stellar insolation values, $S/S_{\oplus}$\,=\,1.0, 0.7, 0.4, while rows correspond to different $\mathrm{CO_2}$ partial pressures, p$\mathrm{CO_2}$\,=\,10\,bar, 1\,bar, $10^{-1}$\,bar, $10^{-2}$\,bar, $10^{-3}$\,bar, $10^{-4}$\,bar. Thick black lines indicate the ground truth $P-T$ profile. The inset shows the retrieved constraints on the surface conditions.}
    \label{fig:PT_post_SNR10_R50}
\end{figure*}

\begin{figure*}[!h]
    \centering
    \includegraphics[width=0.79\linewidth]{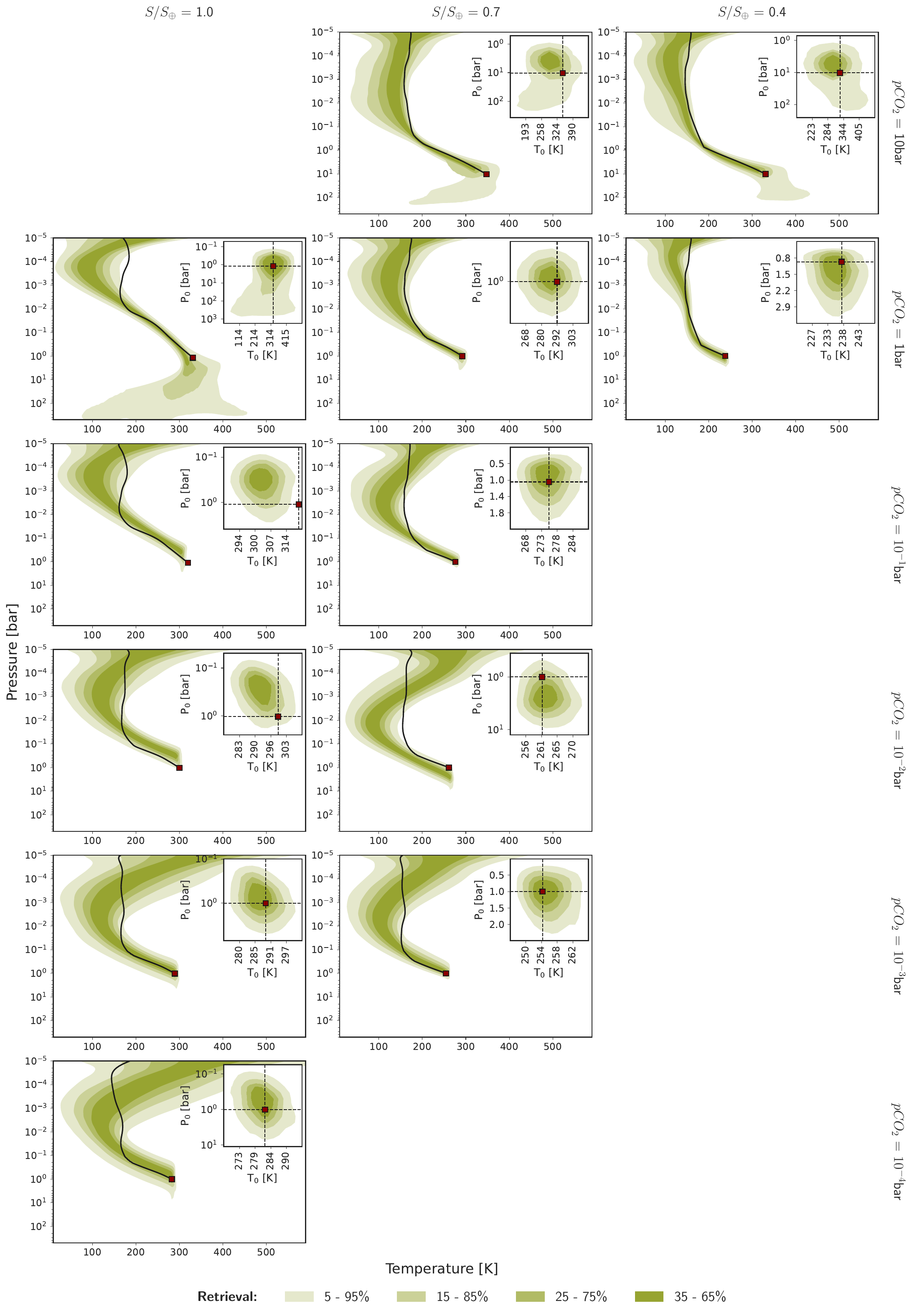}
    \caption{Same as Figure \ref{fig:PT_post_SNR10_R50} but for $S/N$\,=\,10 and $R$\,=\,100.}
    \label{fig:PT_post_SNR10_R100}
\end{figure*}

\begin{figure*}[!h]
    \centering
    \includegraphics[width=0.79\linewidth]{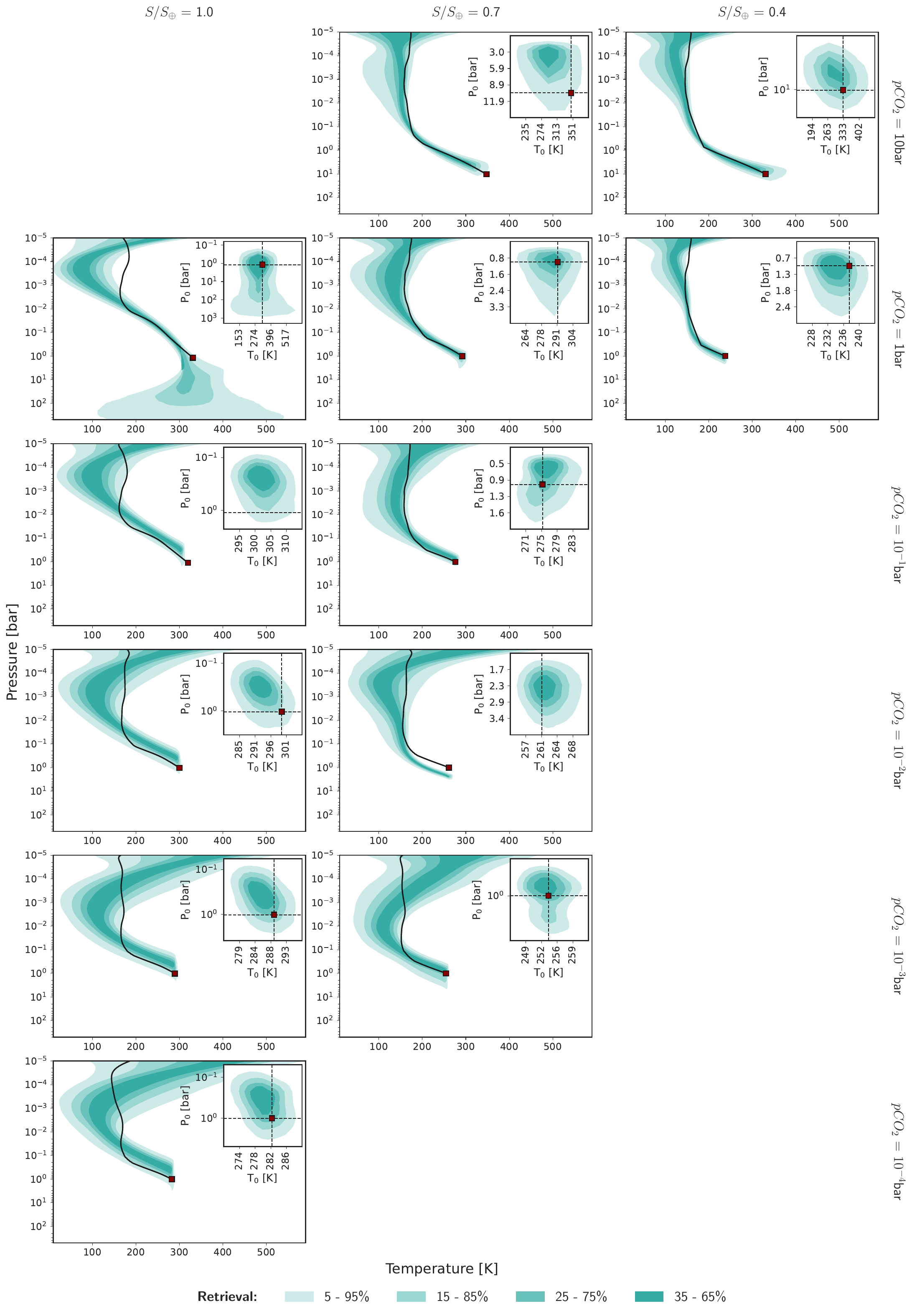}
    \caption{Same as Figure \ref{fig:PT_post_SNR10_R50} but for $S/N$\,=\,20 and $R$\,=\,50.}
    \label{fig:PT_post_SNR20_R50}
\end{figure*}

\begin{figure*}[!h]
    \centering
    \includegraphics[width=0.79\linewidth]{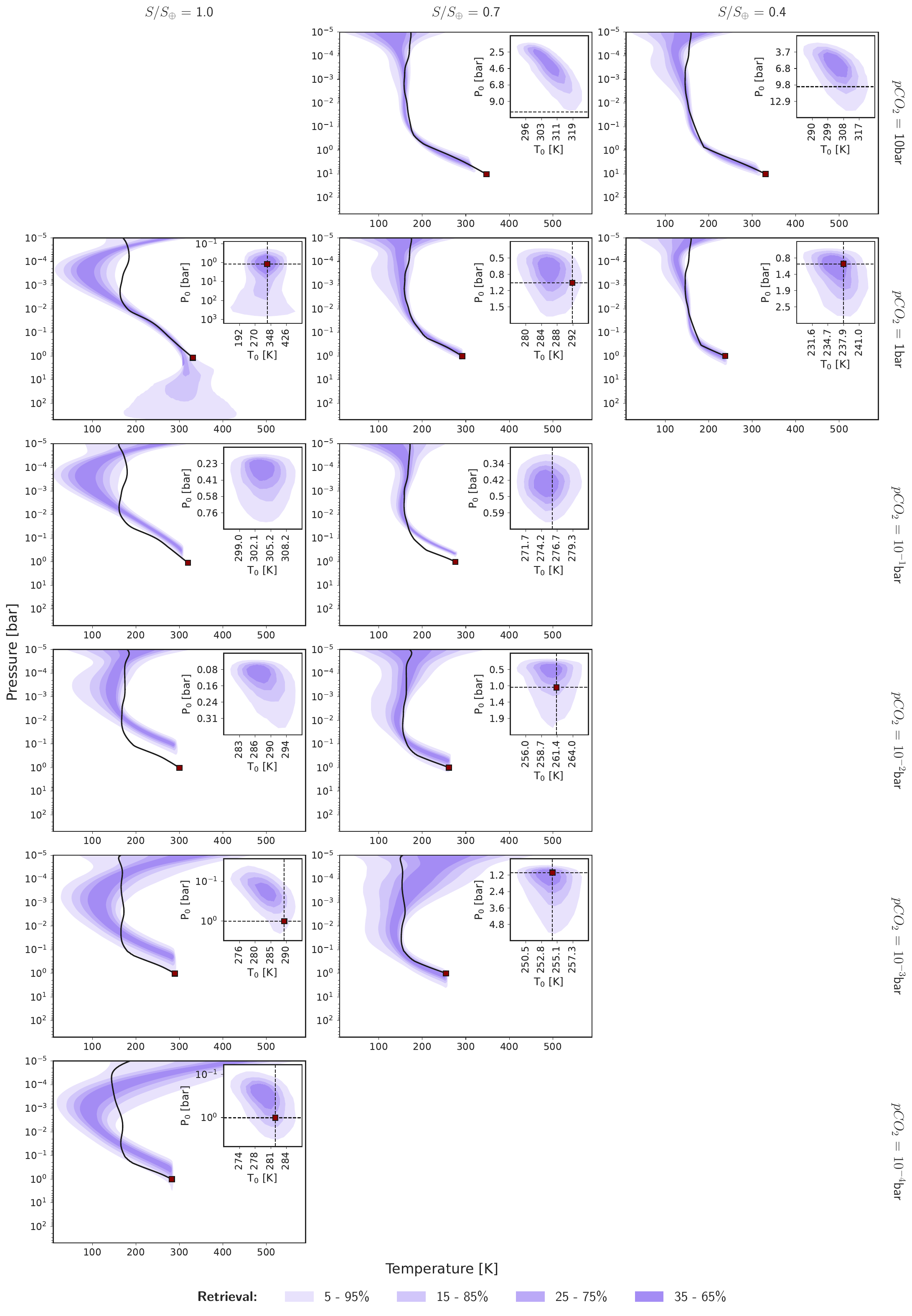}
    \caption{Same as Figure \ref{fig:PT_post_SNR10_R50} but for $S/N$\,=\,20 and $R$\,=\,100.}
    \label{fig:PT_post_SNR20_R100}
\end{figure*}

\begin{figure*}[!h]
    \centering
    \includegraphics[width=0.9\linewidth]{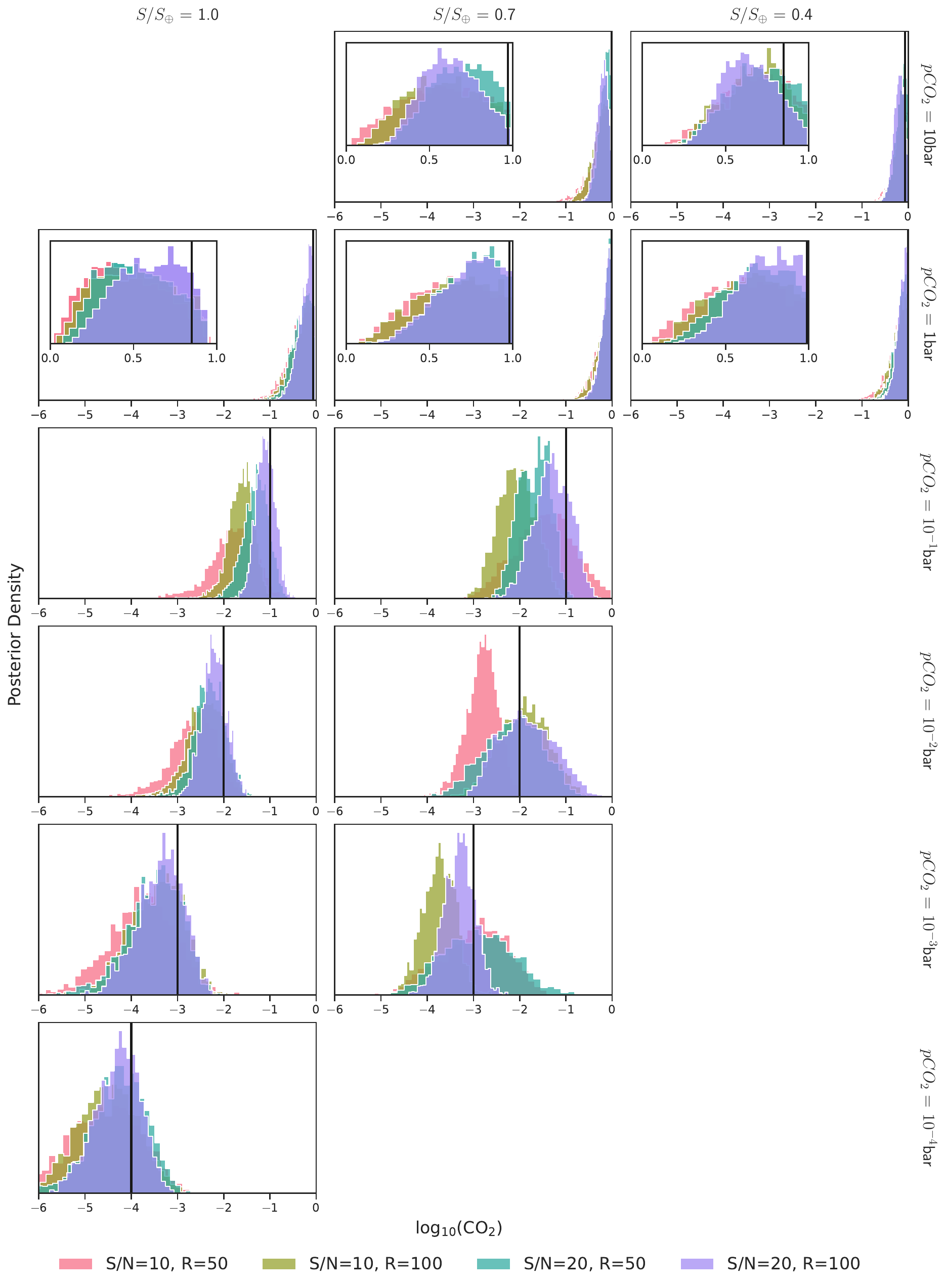}
    \caption{As Figure \ref{fig:CO2_post} but for vertically constant $\mathrm{H_2O}$ profiles in the input spectra.}
    \label{fig:CO2_post_fixedH2O}
\end{figure*}

\begin{figure*}[!h]
    \centering
    \includegraphics[width=0.9\linewidth]{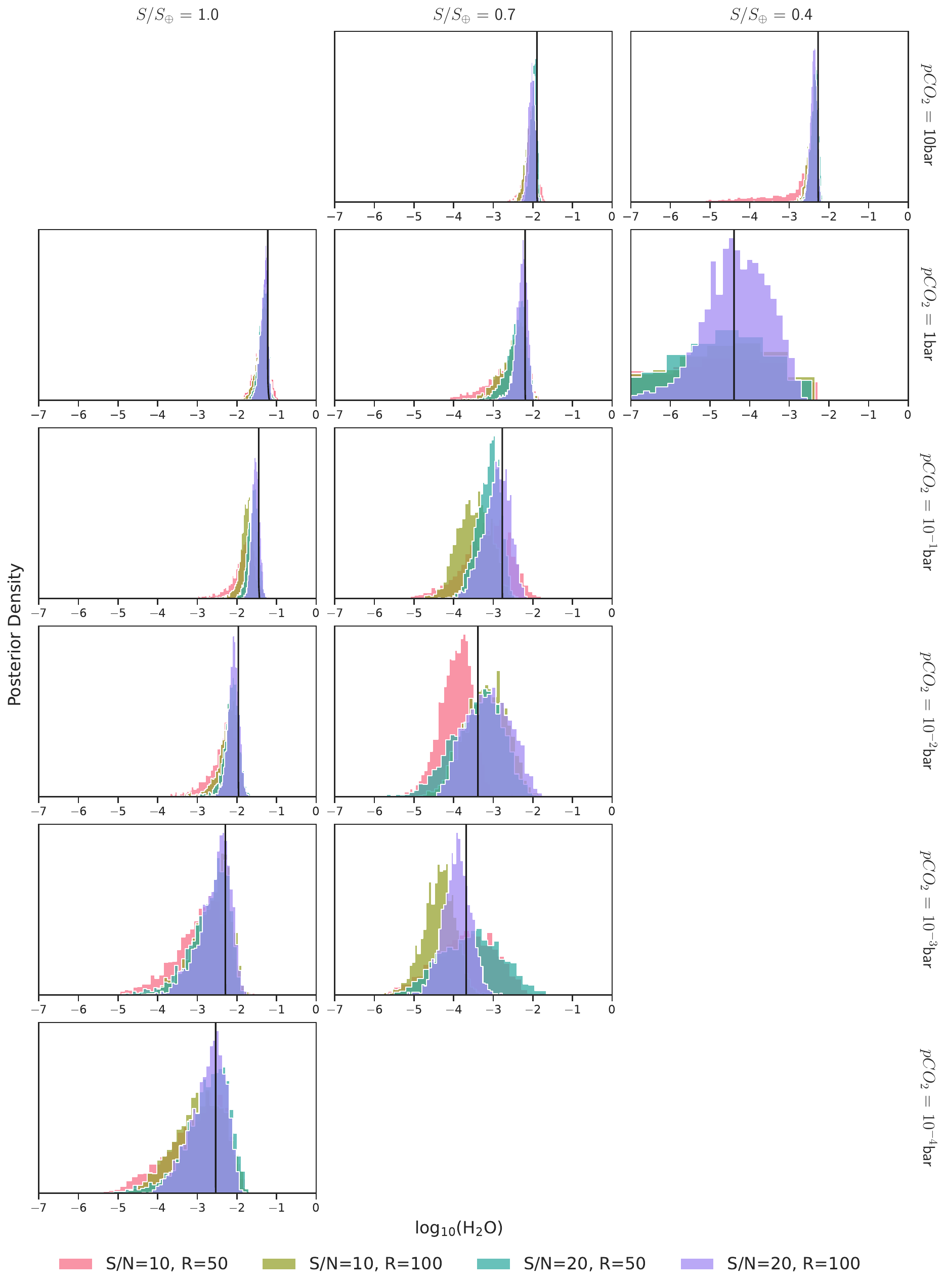}
    \caption{As Figure \ref{fig:H2O_post} but for vertically constant $\mathrm{H_2O}$ profiles in the input spectra. }
    \label{fig:H2O_post_fixedH2O}
\end{figure*}

We show $\sigma$ and $\Delta$ maps for all spectral quality cases in Figures \ref{fig:sigma_map_all} and \ref{fig:delta_map_all}, respectively. We visualize retrieved $\mathrm{CO_2}$ partial pressures for all scenarios in our grid and all spectrum quality cases  covering $S/N$\,=\,[10,\,20] and $R$\,=\,[50,\,100] in Figure \ref{fig:CO2_post_p}. For all spectrum quality scenarios, we show retrieved posterior distributions for $\mathrm{H_2O}$ abundances in Figure \ref{fig:H2O_post} and retrieved $P-T$ profiles in Figures \ref{fig:PT_post_SNR10_R50} - \ref{fig:PT_post_SNR20_R100}. Finally, in Figures \ref{fig:CO2_post_fixedH2O} and \ref{fig:H2O_post_fixedH2O}, we present the $\mathrm{CO_2}$ and $\mathrm{H_2O}$ abundances derived from atmospheric retrievals performed on simulated emission spectra, where $\mathrm{H_2O}$ profiles were assumed to be vertically constant in the input spectra.

\begin{itemize}
    \item Figure \ref{fig:sigma_map_all} - Maps of posterior standard deviations ($\sigma$) for all spectrum quality cases.
    \item Figure \ref{fig:delta_map_all} - Maps of posterior offsets ($\Delta$) for all spectrum quality cases.
    \item Figure \ref{fig:CO2_post_p} - Retrieved $\mathrm{CO_2}$ partial pressure posteriors.
    \item Figure \ref{fig:H2O_post} - Retrieved $\mathrm{H_2O}$ abundance posteriors.
    \item Figure \ref{fig:PT_post_SNR10_R50} - Retrieved $P-T$ profiles for $S/N$\,=\,10 and $R$\,=\,50.
    \item Figure \ref{fig:PT_post_SNR10_R100} - Retrieved $P-T$ profiles for $S/N$\,=\,10 and $R$\,=\,100.
    \item Figure \ref{fig:PT_post_SNR20_R50} - Retrieved $P-T$ profiles for $S/N$\,=\,20 and $R$\,=\,50.
    \item Figure \ref{fig:PT_post_SNR20_R100} - Retrieved $P-T$ profiles for $S/N$\,=\,20 and $R$\,=\,100.
    \item Figure \ref{fig:CO2_post_fixedH2O} - Retrieved $\mathrm{CO_2}$ abundance posteriors for constant $\mathrm{H_2O}$ profiles in input spectra.
    \item Figure \ref{fig:H2O_post_fixedH2O} - Retrieved $\mathrm{H_2O}$ abundance posteriors for constant $\mathrm{H_2O}$ profiles in input spectra.
\end{itemize}

\clearpage
\section{Supplementary Population Trend Retrieval Results}\label{sec:app_supp_trend_retrievals}
\subsection{Population-Trend Retrieval Results for \texorpdfstring{$S/N$}{S/N} = 10 and \texorpdfstring{$R$}{R} = 50}
\begin{figure*}[!h]
    \centering
    \includegraphics[width=\linewidth]{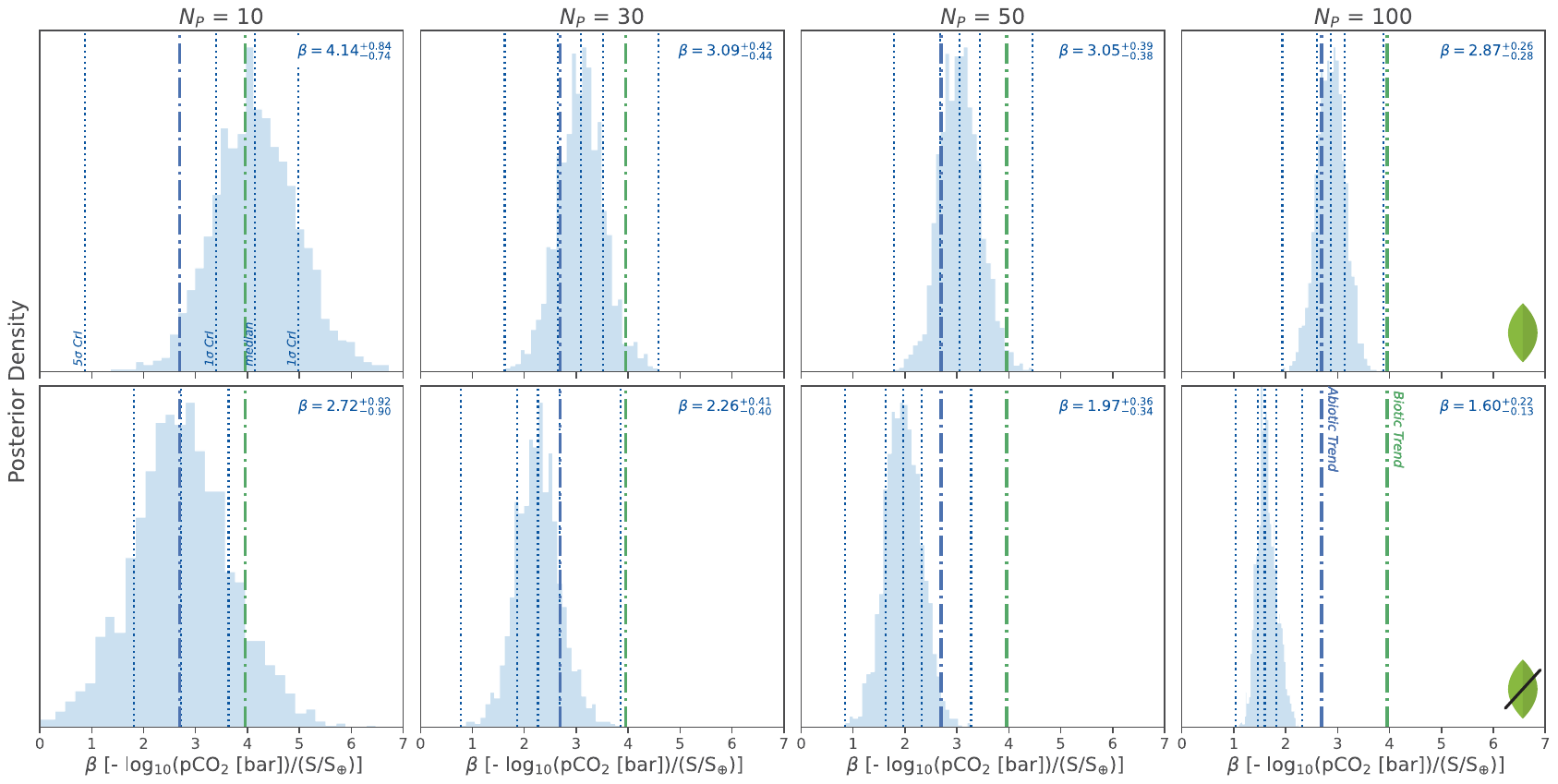}
    \caption{As Figure \ref{fig:HBAR_posterior} but for $S/N$\,=\,10 and $R$\,=\,50.}
    \label{fig:HBAR_posterior_SNR10_R50}
\end{figure*}

We show retrieved posteriors for trend slope parameter $\beta$ for the $S/N$\,=\,10 and $R$\,=\,50 case in Figure \ref{fig:HBAR_posterior_SNR10_R50}, covering $N_P$\,=\,[10,\,30,\,50,\,100] and both biotic and abiotic scenarios.

\subsection{Jeffreys Scale}
We use the Jeffreys scale to quantify model preference and assess the detectability of $\mathrm{CO_2}$ trends by comparing Bayesian evidences, in form of the Bayes' factor $K$, obtained from our trend inference routine with linear and flat trend models. The Jeffreys scale is presented in Table \ref{tab:jeffreys_scale}.

\begin{table}[!h]
{\footnotesize
\caption{Jeffreys scale \citep{Jeffreys_1939} for interpretation of the Bayes' factor $K$.}
\label{tab:jeffreys_scale}
\begin{center}
        \setlength\extrarowheight{4pt} 
\begin{tabular}{lll}
\hline\hline
$\log_{10}(K)$ & Probability & Strength of Evidence \\
\hline
$<$\,0           & $<$\,0.5                              & Support for B    \\
0\,-\,0.5        & 0.5\,-\,0.75                          & Weak support for A    \\
0.5\,-\,1        & 0.75\,-\,0.91                         & Substantial support for A    \\
1\,-\,2          & 0.91\,-\,0.99                         & Strong support for A  \\
$>$\,2           & $>$\,0.99                             & Decisive support for A  \\
\hline
\end{tabular}
\tablecomments{Scale is symmetrical. Hence, negative values of $\log_{10}(K)$ indicate weak, substantial, strong or decisive support for model B.}
\end{center}}

\end{table}

\clearpage
\bibliography{references}{}
\bibliographystyle{aasjournal}

\end{document}